\documentclass
[twocolumn,showpacs,showkeys,preprintnumbers,amsmath,amssymb]{revtex4}

\usepackage{graphicx}
\usepackage{dcolumn}
\usepackage{bm}

\def\bea{\begin{eqnarray}}
\def\eea{\end{eqnarray}}

\begin{document}

\preprint{Version 2.9}

\title{A power-law description of heavy ion collision centrality}

\author{Thomas A. Trainor and Duncan J. Prindle}
\address{CENPA, Box 354290, University of Washington, Seattle, WA 98195}

\date{\today}

\begin{abstract}
The minimum-bias distribution on heavy ion collision multiplicity $n_{ch}$ is well approximated by power-law form $n_{ch}^{-3/4}$, suggesting that a change of variable to $n_{ch}^{1/4}$ may provide more precise access to the structure of the distribution and to A-A collision centrality. We present a detailed centrality study of Hijing-1.37 Monte Carlo data at 200 GeV using the power-law format. We find that the minimum-bias distribution on $n_{participant}^{1/4}$, determined with a Glauber Monte Carlo simulation, is uniform except for a 5\% sinusoidal variation. The power-law format reveals precise linear relations between Glauber parameters $n_{part}$ and $n_{bin}$ and the fractional cross section. The power-law format applied to RHIC data facilitates incorporation of extrapolation constraints on data and Glauber distributions to obtain a ten-fold improvement in centrality accuracy for peripheral collisions.
\end{abstract}



\maketitle

\section{Introduction}

Two-particle correlations in RHIC heavy ion  collisions change rapidly with collision centrality, reflecting strong changes in collision dynamics~\cite{ptprl,axialcd,axialci,ptsca}. Some variations relate to the changing geometry of hadronization~\cite{axialci}. Other variations arise from copious low-$Q^2$ parton scattering~\cite{lowq2}. Correlations from p-p collisions with simpler dynamics provide a precision reference for peripheral Au-Au collisions~\cite{ppmeas}. However, conventional RHIC centrality methods have been ineffective for the 20\% most peripheral collisions, where changes relative to p-p are rapid and  informative. We are motivated therefore to improve centrality determination for A-A collisions.

The {\em power-law} representation of the minimum-bias distribution on collision multiplicity provides a basis for major improvement. The minimum-bias distribution, conventionally plotted in a semi-log format, is approximately a {power-law distribution}~\cite{dave}, implying that a change of variable should lead to a more compact form of the distribution. The ideal power-law form is a uniform distribution between two well-defined endpoints. Deviations from that {precision reference} are easily identified and studied. 

The power-law format helps in several ways. 1) It provides accurate centrality determination down to N-N collisions, even if the measured `minimum-bias' distribution is strongly distorted or biased by triggering and vertex reconstruction inefficiencies. 2) Applied to the Glauber model the power-law format provides compact and precise representations of the Glauber parameters. 3) The power-law plotting format reveals distribution details at the few-percent level important for precision comparisons with p-p collisions.

\section{Method}

The novel techniques described in this paper emerged from the observation that minimum-bias distributions on $n_{part}$ (participant nucleon number), $n_{ch}$, $p_t$ and $E_t$ vary approximately as $x^{-3/4}$, a power law. Transforming the distributions to $x^{1/4}$ has led to major improvements in analysis accuracy. Through running integrals of several power-law distributions we are able to relate measured quantities and geometry parameters at the percent level. The methods are applied to Hijing data and RHIC data.

We first introduce the {power-law} form of the  minimum-bias distribution, compare it to the conventional semi-log form and describe its properties. We then review the Glauber model of A-A collision geometry and study the power-law form of the Glauber minimum-bias distributions on $n_{part}$ and $n_{bin}$. We construct parameterizations of running integrals which relate the Glauber parameters to the fractional cross section. We then demonstrate power-law centrality determination with Hijing and RHIC data, relating $n_{ch}$ to fractional cross section with percent errors. 

To demonstrate the overall method we study the centrality dependence of particle and $p_t$ production in Hijing for two event classes (quench-on, quench-off) and particle and $E_t$ production in RHIC data. Such production studies provide the most demanding test of centrality precision by comparing centralities from data with centralities from a Glauber Monte Carlo. Finally, we make a detailed comparison of systematic centrality errors in conventional and power-law contexts.

We also include three appendices: A) Particle-production algebra: a unified approach to particle production in different plotting contexts; B) Numerical integration techniques: methods of binning, numerical integration and running integrals; C) Relating centrality parameters: what lies behind centrality methods -- joint and marginal distributions, running integrals of marginals, the participant-scaling model and the role of fluctuations.

\section{Hijing Data}

Hijing-1.37~\cite{hijmc} was used to produce minimum-bias event ensembles with $\sim $ 1M total events for each of two classes: 1) {\em quench-off}\, Hijing -- jet production but no jet quenching  and 2) {\em quench-on} Hijing -- jet production with jet quenching. Charged particles with pseudorapidity $|\eta| < $ 1, transverse momentum $p_t \in [0.15,2]$ GeV/c and full azimuth were accepted. The total charged-particle multiplicity and transverse momentum in the acceptance for each event defines centrality parameters $n_{ch}$ and $p_t$. The variations with centrality of Hijing quench-on and quench-off event classes are distinguishable, but to an extent {depending on the analysis method}. We use the two Hijing classes to illustrate the sensitivity of different centrality formats to particle and $p_t$ production mechanisms, as well as the general features of power-law centrality determination.

\section{The Conventional Minimum-bias Distribution}

\begin{figure}[h]
\includegraphics[keepaspectratio,width=3.3in]{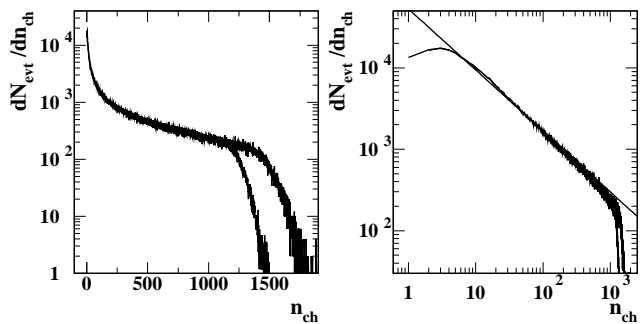}
\caption{Left panel: Minimum-bias distributions for quench-off and quench-on (larger endpoint) Hijing collisions plotted with a conventional semi-log plotting format.  Right panel: The same distributions plotted in a log-log format. The line represents an $n_{ch}^{-3/4}$ power-law trend.  
\label{fig1}}
\end{figure}

In Fig.~\ref{fig1} (left panel) we plot the minimum-bias distribution of event number on multiplicity $n_{ch}$ in a conventional semi-log format for the two Hijing event types. 
The distributions are monotonically decreasing, and the tail widths at large $n_{ch}$ reflect fluctuations in particle production for central collisions ($b = 0$). The fluctuation magnitude (relative to Poisson) depends on the detector acceptance. Comparing the two Hijing event types, the only apparent difference is the end-point positions.

In Fig.~\ref{fig1} (right panel) the same data are plotted in a log-log format which reveals that the data are well approximated by power-law trend $n_{ch}^{-3/4}$ (solid line)~\cite{dave}. The power-law trend suggests that a change of variable could provide more precise access to distribution structure. If $d\sigma / dn_{ch} \propto n_{ch}^{-3/4}$ then $n_{ch}^{3/4}\, d\sigma / dn_{ch} \sim$ constant. Interpreting $n_{ch}^{3/4}$ as the Jacobian of a variable transformation, i.e., $ dn_{ch} = 4 n_{ch}^{3/4}\,dn_{ch}^{1/4}$, we expect $d\sigma / dn_{ch}^{1/4} \sim$ constant. We convert the minimum-bias Hijing data distributions to {\em power-law} form $d\sigma / dn_{ch}^{1/4} \equiv 4 n_{ch}^{3/4}\, d\sigma / dn_{ch}$ and thereby obtain more precise access to data and improved centrality determination.

\section{The Power-law Distribution}

We introduce the power-law minimum-bias distribution and consider examples from Hijing Monte Carlo data and RHIC data. We identify the principal features of the distribution in comparison to a participant-scaling reference. We parameterize the form of the distribution based on a two-component model of particle production.

\subsection{Hijing}

\begin{figure}[h]
\includegraphics[width=1.6in,height=1.7in]{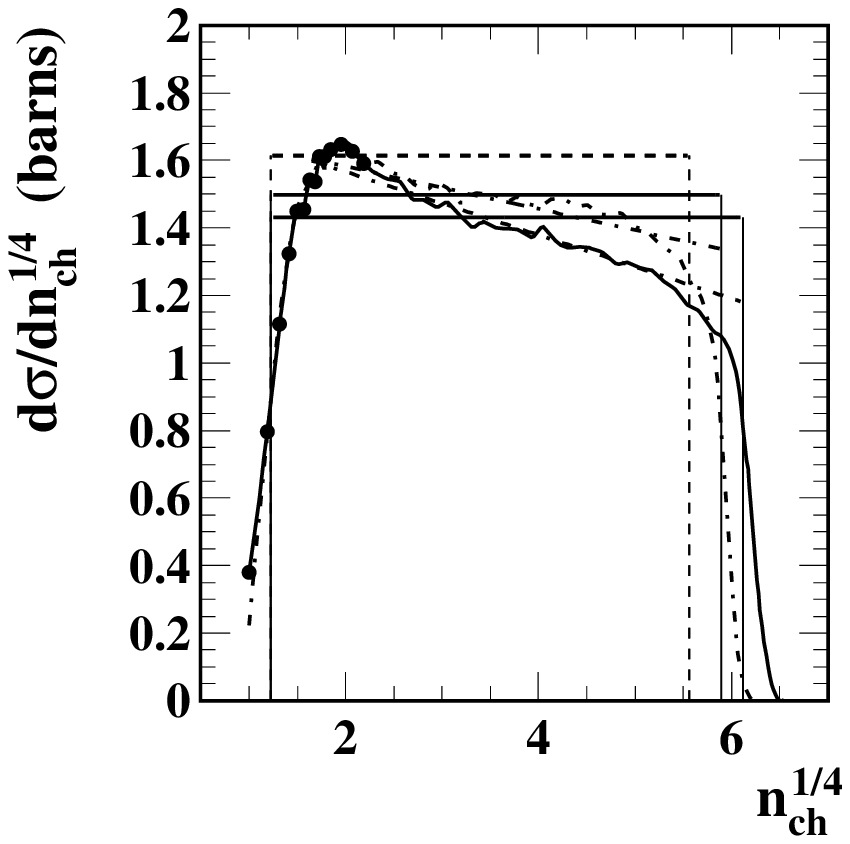}
\includegraphics[width=1.7in,height=1.67in]{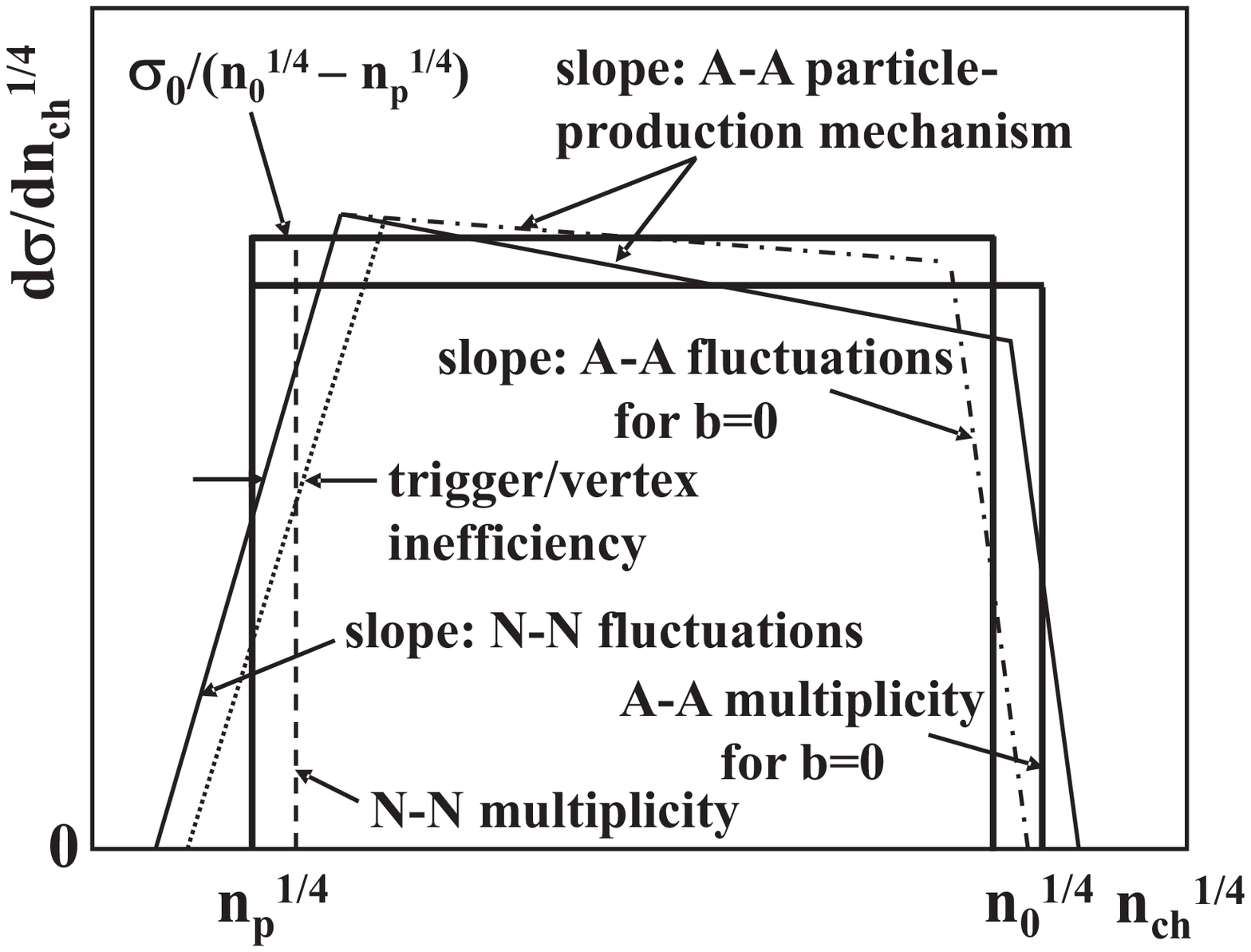}
\caption{Left panel: The power-law form of the minimum-bias distribution for quench-on (solid curve) and quench-off (dash-dot curve) Hijing data. The solid rectangles are power-law references for the data. The dashed rectangle represents participant scaling. The dash-dot lines are as defined for Fig.~\ref{fig3} (right panel). Right panel: A cartoon of the power-law minimum-bias distribution identifying the principal features. The vertical solid lines indicate the positions of half-maximum endpoints $n_p^{1/4}$ and $n_0^{1/4}$. The dashed line indicates the position for the N-N multiplicity $n_{NN} \sim 2\, n_{p}$. 
\label{fig2}
}
\end{figure}

Fig.~\ref{fig2} (left panel) shows distributions $d\sigma / dn_{ch}^{1/4}$ {\em vs} $n_{ch}^{1/4}$ for two Hijing configurations which confirm the basic features anticipated for the power-law format---an approximately rectangular distribution with limited amplitude variation and well-defined endpoints. The distribution on integers has been rebinned to 50 uniform bins on $n_{ch}^{1/4}$ ({\em cf.} App.~\ref{rebinning}), insuring nearly-uniform statistical bin errors while retaining adequate resolution near the upper endpoint. The solid points at the lower endpoint indicate the limiting edge resolution. The first few points are defined by the smallest integers, not the rebinning. The lower endpoint (half-maximum point) is $n_p \sim n_{NN}/2$ (but {\em cf.} App.~\ref{intfluct}). The upper endpoints $n_0 =$1210 and 1400 are not exactly at the half-maximum points due to the asymmetric (skewed) shape of fluctuations for central collisions as modeled by Hijing.  The sloped dash-dot lines starting at $n_{NN}^{1/4}$ and approximating the data are defined in the next subsection.

The five parameters which describe the shape of the power-law minimum-bias distribution are summarized in Fig.~\ref{fig2} (right panel): 1) Upper half-maximum endpoint $n_0$ estimates the mean $n_{ch}$ corresponding to central A-A collisions ($b = 0$) and upper endpoint $n_{part,0}/2$ of the minimum-bias distribution on participant-pair number $n_{part}/2$ ({\em cf.} Sec.~\ref{powerglaub} for $n_{part}$ endpoint definitions). 2) Lower half-maximum endpoint $n_p$ is approximately one-half the mean $n_{ch}$ for N-N ($\sim$ p-p) collisions (lower endpoint $n_{part,p}/2 \sim 1/2$). 3), 4)~The slopes at the endpoints measure particle-production fluctuations in N-N and $b = 0$ A-A collisions, which also depend on the detector acceptance~\cite{relflucts}.  5) The slope near the midpoint reflects the particle-production mechanism, i.e., the relative importance of binary-collision and participant-pair scaling. At 200 GeV the Hijing (Pythia) N-N multiplicity is $n_{NN} \sim 5$ in $|\eta| < 1$. For a symmetric N-N multiplicity distribution the lower half-maximum point would be $n_p =  n_{NN}/2$. However, for Hijing (and data)  $n_p \sim 0.45\, n_{NN}$ because the N-N distribution is significantly skewed (0.45 estimates $n_{part,p}/2$, {\em cf.} App.~\ref{intfluct}).

The power-law {\em reference distribution} is a rectangle (uniform distribution on $n_{ch}^{1/4}$) bounded by endpoints $n_p^{1/4}$ and $n_0^{1/4}$, with area $\sigma_0$ the total cross section defined by the event trigger. The average value of the power-law minimum-bias distribution is therefore $\sigma_0 / (n_0^{1/4} - n_p^{1/4})$.  The solid rectangles in Fig.~\ref{fig2} (left panel) are power-law references for Hijing data, with upper endpoints  $n_0 = 1210 ~(1400)$ for quench-off (quench-on) events within two units of pseudorapidity.  The dashed rectangle is a {\em participant scaling} reference with lower endpoint  $n_p = n_{part,p}/2\cdot n_{NN}$ (in common with the solid rectangles) and upper endpoint $n_0 \rightarrow n_0^* = n_{part,0}/2 \cdot n_{NN}$. Systematic deviations from the ideal power-law distribution near $n_p$ could result from trigger and/or vertex inefficiencies, or contamination from non-hadronic backgrounds. The five shape parameters provide precise determination of collision centrality and particle (or $p_t$, $E_t$) production. More detailed features are suppressed in the running integrals used to relate $n_{ch}$, $p_t$ and $E_t$ to fractional cross section $\sigma / \sigma_0$.

Fig.~\ref{fig2} (left panel) demonstrates that the power-law format clearly reveals physics-related differences  between quench-on and quench-off distributions at the few-percent level not apparent in the semi-log plotting format of Fig.~\ref{fig1}. We now apply the power-law format to a minimum-bias distribution obtained from RHIC data. 

\subsection{RHIC data}

\begin{figure}[h]
\includegraphics[width=3.3in,height=1.7in]{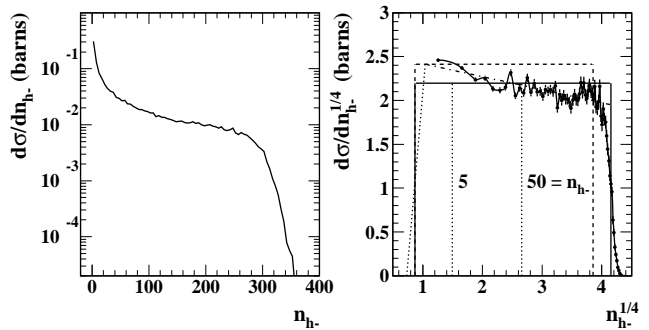}
\caption{Left panel: A minimum-bias distribution of RHIC data~\cite{spectra1} with a conventional semi-log plotting format. Right panel: The same data plotted with a  power-law format. The solid rectangle is a 7.2 barn power-law reference for the data. The dashed rectangle is the corresponding participant-scaling reference. The sloped dash-dot line is inferred from endpoint $n_0$ of the data and the mean NSD p-p multiplicity. 
\label{fig3}}
\end{figure}

Fig.~\ref{fig3} shows a minimum-bias distribution on $n_{h^-}$ (negative hadron multiplicity) in $|\eta|<0.5$ for 60k Au-Au collisions at $\sqrt{s_{NN}} =$ 130 GeV~\cite{spectra1}. The distribution was corrected for trigger and tracking inefficiencies and backgrounds. Its analysis illustrates some aspects of the collision geometry problem relating to real data. The left panel shows the conventional semi-log plotting format with uniform bins on $n_{h^-}$. The assumed total cross section is $\sigma_{0} =$ 7.2 barns~\cite{baltz}. The event-trigger efficiency (coincidence of two ZDCs~\cite{spectra1}) was greater than 98\% for all multiplicities. 

The measured event-vertex efficiency was 100\% for $n_{h^-} > 50$ but dropped to 60\% below $n_{h^-} = 5$. The contribution to $\sigma_{0}$ from the lowest bin was estimated to be 21\% based on an extrapolation using Hijing. It was argued that peripheral A-A collisions are linear superpositions of N-N collisions, and Hijing (Pythia) models N-N collisions correctly. Hijing was normalized to data for $n_{h^-} \in [5,25]$  ($n_{h^-}^{1/4} \in [1.5,2.24]$) and used for the extrapolation below $n_{h^-} = 5$. The estimated systematic uncertainty in the inferred differential cross section was 10\%, due to uncertainties in $\sigma_{0}$ and the inferred relative contribution from the first bin based on the Hijing extrapolation. 

Fig.~\ref{fig3} (right panel) shows the same data plotted in the power-law format (including the first bin of the analysis, bounded above by the dotted line at $n_{h^-} = 5$). The figure confirms that the power-law format is also applicable to RHIC data. Quantitative details revealed in the linear power-law format are not accessible in the semi-log format of the left panel. Two of the five power-law parameters can be obtained directly from the data: the upper endpoint $n_0^{1/4} \sim 4.15 = 297^{1/4}$ (mean $n_{h^-}$ for $b=0$) and the fluctuation width (slope) at $n_0$. The shape about the upper endpoint indicates that fluctuations are nearly symmetric about $n_0$, in contrast to the significant skewness of Hijing (Fig.~\ref{fig2} -- left panel). Lower endpoint $n_p$ and the fluctuation slope  at $n_p$ are not directly accessible due to efficiency and background uncertainties and the binning scheme (linear on $n_{h^-}$). However, the left edge can be sketched (dotted line at left endpoint) based on the expected p-p $h^-$ yield. The corrected $n_{NN} = n_{NN,h^+} + n_{NN,h^-}$ at 130 GeV should be $\sim$2.3 in one unit of rapidity~\cite{fabe}. Thus, we estimate $n_{p}^{1/4} \approx (0.45\, n_{NN,h^-})^{1/4} =0.85$. 

The solid rectangle represents the {\em power-law reference} corresponding to $\sigma_{0} =$ 7.2 barns distributed uniformly on the interval between the estimated $n_p^{1/4}$ and the observed $n_0^{1/4}$. The dashed rectangle represents the {\em participant-scaling} reference, where $n_0^{1/4} \rightarrow (n_0^*)^{1/4} \equiv (n_{NN,h^-} \cdot n_{part,0}/2)^{1/4} = (1.15 \cdot 191)^{1/4} = 3.85 $. 
Combining the two references we can predict the average slope of the data distribution between endpoints. Since the total cross section is the same for both data and participant-scaling reference the slope is $m = - 2A(n_0^{1/4} - [n_0^*]^{1/4}) / (n_0^{1/4} - [n_{NN,h^-}]^{1/4})^2 = -0.15$, where $A = d\sigma/dn_{h^-}^{1/4} \rightarrow  2.4$ barns is the uniform differential cross section for participant scaling, with $n_0^{1/4}$, $n_{NN,h^-}$ and $(n_0^*)^{1/4}$ evaluated above. The dash-dot line drawn with that slope starting at $(n_{NN,h^-})^{1/4}$ forms a trapezoid approximation to the data. The fractional change in height is $0.15\times 3.3/2.2  \approx $ 22\%, consistent with two-component parameter $x = 0.08$, as shown in App.~\ref{partprod2}.

The comparison with data indicates good agreement between the results of~\cite{spectra1} and the power-law description with N-N (p-p) constraint. The two equivalent physics results from the analysis are the negative slope of the power-law distribution and the difference between $n_0$ values for data and the participant-scaling reference ({\em cf.} App.~\ref{nparscale}). In essence, given endpoint $n_0$ and the NSD N-N (p-p) multiplicity the entire minimum-bias distribution is known sufficiently well for centrality determination at the percent level. This review of a conventional centrality analysis in a power-law context provides some idea of the precision possible with the power-law format.

\section{The Glauber Model}

The Glauber model of nucleus-nucleus collisions represents the multiple nucleon-nucleon interactions within the two-nucleus overlap region in a simply calculable form. The nuclear-matter distribution is modeled by a Woods-Saxon (W-S) function. The nucleon distribution can be modeled as a continuum W-S distribution (so-called optical Glauber) or as a random nucleon distribution sampled from the W-S density (Monte Carlo Glauber). The Glauber geometry parameters are participant number $n_{part}$, N-N binary-collision number $n_{bin}$, nucleon mean path length $\nu = 2\, n_{bin} / n_{part}$ and A-A cross section $\sigma(b)$. Precise determination of $n_{part}$, $n_{bin}$ and $\nu$ in relation to impact parameter $b$ and fractional cross section $\sigma(b) / \sigma_{0}$ establishes the centrality dependence of particle production and correlations. The following descriptions of optical and Monte Carlo procedures are derived in part from~\cite{miller,starglaub}.
 
\subsection{Optical Glauber}

The optical Glauber relates A-A geometry parameters to $b$ through continuous integrals of the nuclear density. Normalized function $\rho_A(\vec{r})$ is a 3D nuclear density with a Woods-Saxon radial form. The projection onto a plane normal to collision axis $z$ (single-particle areal density) is defined by $T_A(\vec{s}) \equiv \int dz\,\rho_A(\vec{s},z)$. The {\em overlap integral} (two-particle areal density) for nuclei $A$ and $B$ is $T_{AB}(b) \equiv \int d\vec{s}\,\,T_A(\vec{s})\,T_B(\vec{s}-\vec{b})$, an autocorrelation distribution for $T_A(\vec{s})$ if $A = B$. $n_{part}(b)$, $n_{bin}(b)$ and $d\sigma(b)/d\pi b^2$ are defined as integrals of $T_A$ combined with the appropriate nucleon-nucleon cross section $\sigma_{NN}$~\cite{miller}. $n_{bin}(b) \equiv \int d\vec{s}\, A T_A(\vec{s})\, \sigma_{NN}\, B T_B(\vec{s} - \vec{b})$ and for $A,B \gg 1$, $d\sigma(b)/d\pi b^2 \simeq 1 - \exp\{- n_{bin}(b)\}$ and $n_{part}(b)/2 \simeq \int d\vec{s}\, A T_A(\vec{s})\, \{1 - \exp[- \sigma_{NN}\, B T_B(\vec{s} - \vec{b})] \}$. The expression for $d\sigma(b)/d\pi b^2$ implies that $n_{bin}(b_0) \sim \ln(2) < 1$ (and actually goes to zero for large $b$), whereas we expect $n_{bin} \rightarrow 1$ {\em from above} for peripheral collisions of real nuclei. The same problem arises for $n_{part}$, since $n_{part}/2 \rightarrow n_{bin}$ for large $b$.

\subsection{Monte Carlo Glauber}

The Monte Carlo Glauber simulates an ensemble of \mbox{A-B} nucleus-nucleus collisions for a distribution of impact parameters $b$ uniform on $b^2$. Each simulated collision combines discrete nucleon distributions sampled randomly from the continuous Woods-Saxon nuclear densities $\rho_A(\vec{r})$, $\rho_B(\vec{r})$. From the event ensemble $n_{part}(b)$ and $n_{bin}(b)$ are sampled as correlated random variables. Minimum-bias differential cross-section distributions on $n_{part}$ and $n_{bin}$ are constructed from those data. 

\begin{figure}[h]
\includegraphics[keepaspectratio,width=3.3in]{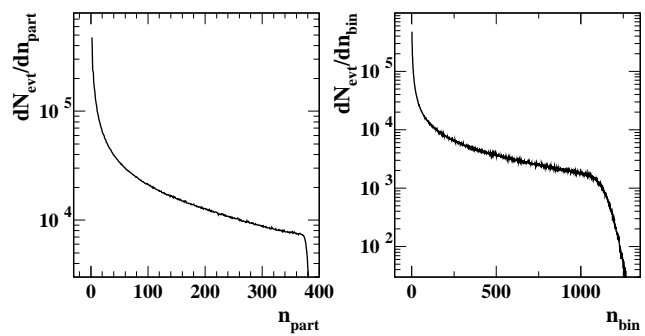}
\caption{Left panel: The conventional minimum-bias distribution on participant number obtained from a Monte Carlo Glauber simulation~\cite{gonz}. Right panel: The corresponding distribution on binary-collision number. 
\label{fig4}}
\end{figure}

Fig.~\ref{fig4} shows minimum-bias distributions from a Monte Carlo Glauber simulation for 200 GeV Au-Au collisions~\cite{gonz} plotted in the conventional semi-log format. When plotted in a log-log format the distribution on $n_{part}$ closely follows the power-law trend $n_{part}^{-3/4}$ similar to data, whereas the distribution on $n_{bin}$ follows the power-law trend $n_{bin}^{-5/6}$. Those trends suggest transformations to power-law distributions on $n_{part}^{1/4}$ and $n_{bin}^{1/6}$. 

\subsection{Power-law Glauber} \label{powerglaub}

\begin{figure}[h]
\includegraphics[width=1.65in,height=1.65in]{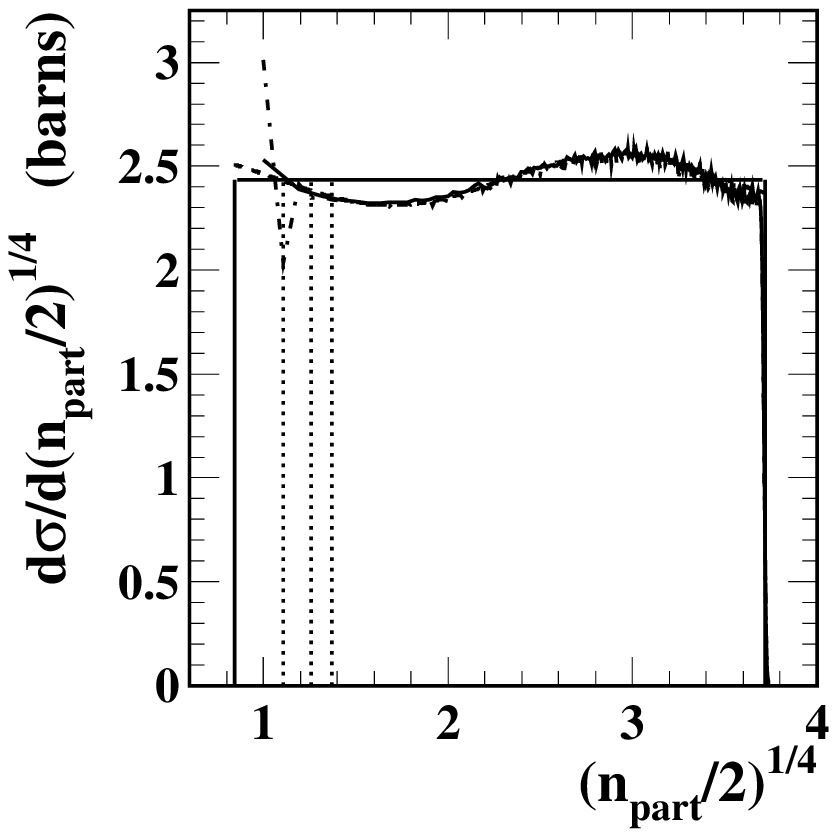}
\includegraphics[width=1.65in,height=1.65in]{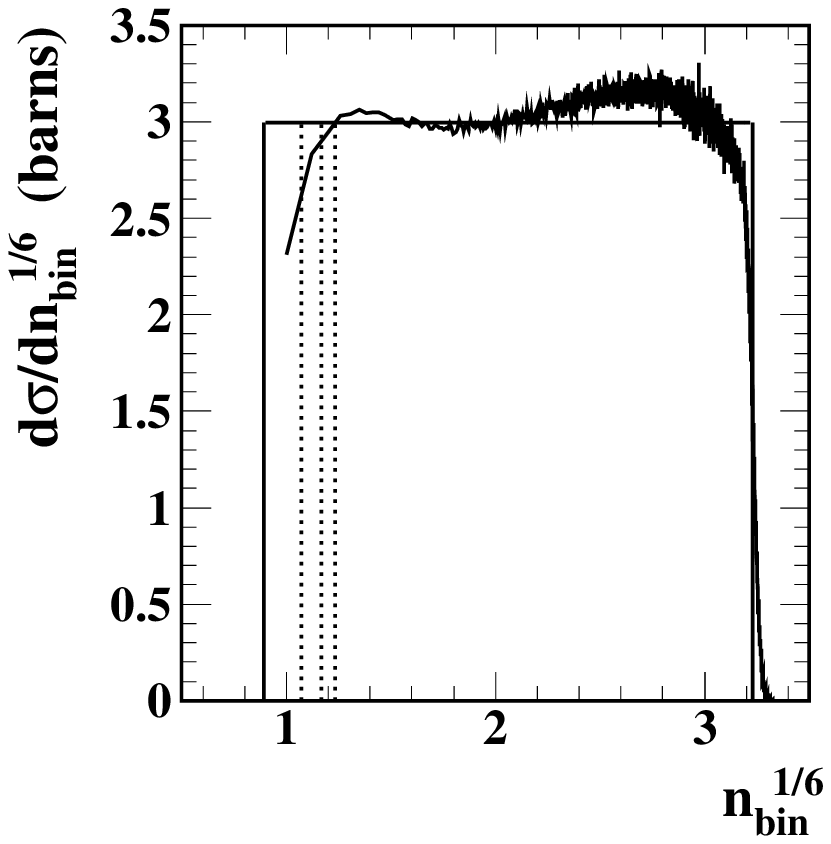}
\caption{Power-law minimum-bias distributions on participant-pair (left panel) and binary-collision (right panel) numbers, obtained from a Glauber Monte Carlo simulation~\cite{gonz}. The assumed total cross section is $\sigma_0 = 7$ barns. The upper endpoints (half-maximum points) for $n_{part}^{1/4}$ and $n_{bin}^{1/6}$ are 3.72 and 3.23 respectively, and the lower endpoints are $0.5^{1/4}$ and $0.5^{1/6}$ respectively. 
\label{fig5a}}
\end{figure}

In Fig.~\ref{fig5a}  we plot power-law minimum-bias distributions on $(n_{part}/2)^{1/4}$ and $n_{bin}^{1/6}$ from the Monte Carlo Glauber data in Fig.~\ref{fig4}. The distributions are nearly rectangular, and bounded on the right end by endpoints $n_{part,0}/2 = 191$ and $n_{bin,0} = 1136$, with $(n_{part,0}/2)^{1/4} = 3.72$ and $n_{bin,0}^{1/6} = 3.23$. The lower endpoints $n_{part,p}/2$, $n_{bin,p}$ and binning scheme (dotted  lines) are discussed in the next section. The distribution on $(n_{part}/2)^{1/4}$ is especially simple: a constant plus a sinusoid with 5\% relative amplitude described by
\bea
\frac{d\sigma}{d(n_{part}/2)^{1/4}} &&= \frac{\sigma_0}{n_{part,0}^{1/4} - n_{part,p}^{1/4}} \times \\ \nonumber
 && \hspace{-.2in} \left\{ 1+ 0.05 \sin\left(2.6\left[(n_{part}/2)^{1/4}-2.3\right]\right) \right\}
\eea
with centroid $2.3 = \{(n_{part,0}/2)^{1/4} + (n_{part,p}/2)^{1/4}\}/2 = (191^{1/4} + 0.5^{1/4})/2$. That expression is plotted as the dashed curve just visible at the left end of the left panel.

In App.~\ref{runbin} we discuss several problems in relating Glauber parameters to the fractional cross section with these differential distributions. A critical issue is how the integer spaces should be binned and integrated. For the integration method used to obtain dash-dot curves in Figs.~\ref{fig5b} and \ref{fig5c} the dotted lines in Fig.~\ref{fig5a} represent the first three bins. The dash-dot curve in the left panel passes through the original Monte Carlo data distribution on $n_{part}$. The solid curve passes through a regrouped distribution on integer $n_{part}/2$ consistent with assumptions in the participant-scaling hypothesis (App.~\ref{nparscale}). 

The power-law format reveals details at the percent level inaccessible with the semi-log format of Fig.~\ref{fig4}. The observed {\em approximate} power-law trend for the minimum-bias distribution on $n_{ch}$ is a consequence of the {\em nearly-exact} power-law trend on $n_{part}$. Whatever its origins, we capitalize on the simplicity of the power-law trend to refine centrality measurement and better understand the mechanisms of particle, $p_t$ and $E_t$ production.

\section{Glauber parameterizations}

Figs.~\ref{fig5b} and \ref{fig5c}  show running integrals of Monte Carlo Glauber data which connect $n_{part}/2$ and $n_{bin}$ to centrality measured by the fractional cross section in the form $1 - \sigma / \sigma_0$. For the purpose of centrality determination (to $\sim$ 2\%) the power-law Glauber curves at 200 GeV are well represented by simple linear expressions $(n_{part}/2)^{1/4} =  \sigma / \sigma_0 \cdot (n_{part,p}/2)^{1/4} + (1-\sigma / \sigma_0) \cdot (n_{part,0}/2)^{1/4}$ and   $n_{bin}^{1/6} =   \sigma / \sigma_0 \cdot n_{bin,p}^{1/6} + (1-\sigma / \sigma_0) \cdot n_{bin,0}^{1/6}$. The dotted lines in Figs.~\ref{fig5b} and \ref{fig5c} represent those {\em power-law references}, with lower endpoints $n_{part,p}/2$ and $n_{bin,p}$ set equal to 3/4 (upper dotted lines) and 1/2 (lower dotted lines). The correct endpoint choice (1/2) is justified in  App.~\ref{running}. The dashed and dash-dot curves are {running integrals} of the distributions in Fig.~\ref{fig5a} ({\em cf.} App.~\ref{runbin}). Within the power-law format we parameterize the dash-dot curves simply and precisely as the solid curves.  The insets provide details of the peripheral regions. 

\subsection{Running-integral definitions}

\begin{figure}[t]
\includegraphics[width=3.3in,height=3.in]{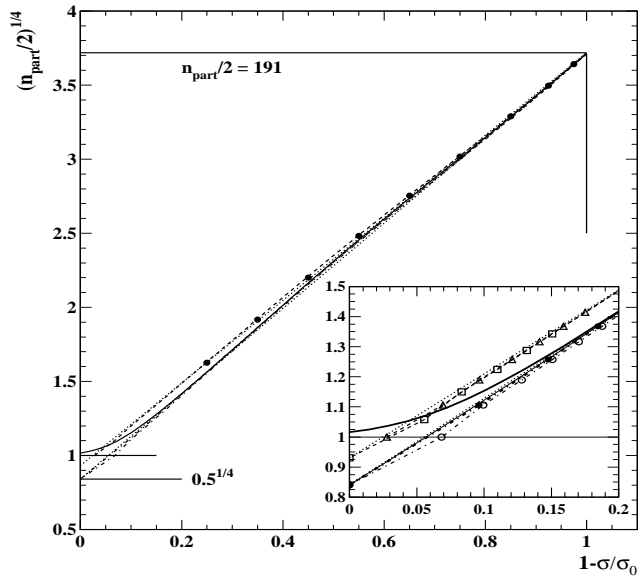}
\caption{Participant-pair number $(n_{part}/2)^{1/4}$ {\em vs} fractional cross section $1 - \sigma/\sigma_0$ in the power-law format. The dotted lines are power-law references for two choices of lower endpoint ($0.5^{1/4}$ and $0.75^{1/4}$). The dash-dot curves are running integrals of the distributions in Fig.~\ref{fig5a} (left panel, solid and dash-dot curves) with endpoints $n_{part,p}/2 = 1/2$. The dashed curves represent alternative running integrals with lower endpoint $n_{part,p}/2 = 3/4$. The solid curves are the  $n_{part}/2$ parameterization described in this section. The solid dots are from a published Monte Carlo Glauber simulation~\cite{starglaub}. 
\label{fig5b}
}
\end{figure}

\begin{figure}[t]
\includegraphics[width=3.3in,height=3.in]{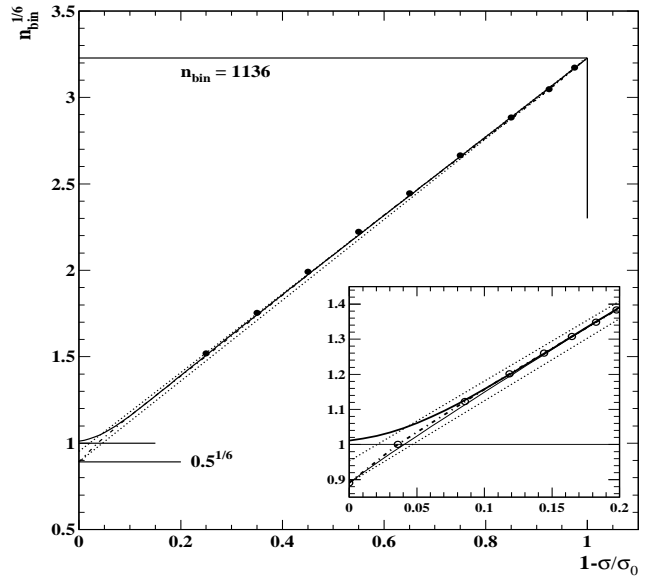}
\caption{N-N binary collision number $n_{bin}^{1/6}$ {\em vs} $1 - \sigma/\sigma_0$. The dotted lines are power-law references for two choices of lower endpoint ($0.5^{1/6}$ and $0.75^{1/6}$). The dash-dot curve is a running integral of the distribution in Fig.~\ref{fig5a} (right panel) with lower endpoint $n_{bin}^{1/6} = (1/2)^{1/6}$. The open circles in the inset are the first few points on the dash-dot curve. The solid curves are the  $n_{bin}$ parameterization described in this section. The solid points are from a Monte Carlo Glauber simulation with binned centrality~\cite{starglaub}.
\label{fig5c}
}
\end{figure}

In Fig.~\ref{fig5b} four forms of running integral (two dashed and two dash-dot curves) are plotted.  The solid curve is the parameterization defined in Eq. (\ref{npart}). The dashed curves, consistent with the power-law reference with endpoint $n_{part,p}/2 = 3/4$ (upper dotted line), can be obtained in two ways ($f_i$ is the $i^{th}$ element of a minimum-bias differential histogram in Fig.~\ref{fig4}, with $M$ elements): 

1) Unit bins on continuous $n_{part}$ are centered on integer $n_{part}$ values. The running sum of $f_i$ is plotted at upper bin edges on $(n_{part}/2)^{1/4}$, defining a {\em middle} Riemann sum on $n_{part}$. The running sum is $F_m \equiv \sum_{i=1}^m f_i $, and the normalized running sum is $1-\sigma/\sigma_0 = F_m / F_M $. Its $m^{th}$ value is plotted at upper bin edge $\{(n_{part,m} + 1/2)/2\}^{1/4}$ (open squares in Fig.~\ref{fig5b} -- inset). The lower endpoint (lowest bin edge) on $n_{part}/2$ is 3/4 ($n_{part,min} = 2$), and the bin edges continue upward on odd quarters. 

2) Unit bins lie between integers on continuous $n_{part}$.  At the $i^{th}$ integer value of $n_{part}$ the differential cross-section entry at that and the proceeding integer $n_{part}$ value are averaged, defining an {\em upper} Riemann sum. The running sum is then $F_m = \sum_{i=2}^m (f_i + f_{i-1})/2$, plotted at upper bin edge $(n_{part,m}/2)^{1/4}$. (open triangles in Fig.~\ref{fig5b} -- inset) The difference between 1) and 2) is mainly that points on $(n_{part}/2)^{1/4}$ are shifted by 1/2 bin. The lower endpoint is also 3/4.  Either method emulates bin averages on $n_{part}$ used to relate $n_{part}$ to $\sigma/\sigma_0$.

The dash-dot curves, consistent with the power-law reference with endpoint $n_{part,p}/2 = 1/2$ (lower dotted line) can also be obtained in two ways ($f_i^{1/4}$ is the $i^{th}$ element of the power-law minimum-bias differential histogram derived from $f_i$, with $M$ elements): 

1)  Unit bins lie {\em between} integers on continuous $n_{part}$. Each entry from the power-law histogram on $n_{part}^{1/4}$ is multiplied by the bin width on $(n_{part}/2)^{1/4}$ {\em preceding} it. The running sum is plotted at the $(n_{part}/2)^{1/4}$ value of the entry, defining an {\em upper} Riemann sum on $n_{part}$.  The running sum is $F_m = \sum_{i=2}^m f_i^{1/4} \cdot [(n_{part,i}/2)^{1/4} - (n_{part,i-1}/2)^{1/4}]$, and the corresponding normalized running sum is plotted at $(n_{part,m}/2)^{1/4}$ (open circles in Fig.~\ref{fig5b} -- inset). The effective lower endpoint (lowest bin edge) on $n_{part}/2$ is 1/2.

2) Unit bins on continuous $n_{part}/2$ are centered on {\em integer} values of $n_{part}/2$. The first few bins are illustrated by the dotted lines in Fig.~\ref{fig5a} (left panel). The $M$ entries of the $f^{1/4}_i$ power-law histogram on $n_{part}^{1/4}$ transformed from $f_i$ on $n_{part}$ (dash-dot curve in that panel) are combined in pairs to form $M/2$ entries $g_j^{1/4}$ on {integer} values of $n_{part}/2$ (solid curve in that panel), with $g_j^{1/4} = f_{2j}^{1/4} + f_{2j+1}^{1/4}$. The corresponding bin widths on $(n_{part}/2)^{1/4}$ are determined exactly, and the running sum is plotted at upper bin edges on $n_{part}/2$, defining a {\em middle} Riemann sum on $n_{part}/2$.  The running sum is $G_m = \sum_{j=1}^m g_j^{1/4}  \cdot [(n_{part,j}/2 + 1/2)^{1/4} - (n_{part,j}/2 - 1/2)^{1/4}]$, and $G_m / G_{M/2}$ is plotted at $(n_{part,m}/2 + 1/2)^{1/4}$ (solid dots in Fig.~\ref{fig5b} -- inset). The effective lower endpoint (lowest bin edge) on $n_{part}/2$ is again 1/2. This last method complies fully with the participant-scaling hypothesis ({\em cf.} App.~\ref{nparscale}). Bin averages on $n_{part}/2$ should be consistent with these results.

\subsection{Full power-law parameterizations}

The simple linear power-law parameterizations described in the beginning of this section provide good visual comparisons and specific expectations for limiting cases (extrapolation constraints). For {\em particle production} studies the accuracy demand on $n_{part}/2$ increases substantially, and the 5\% sinusoid should be included in the $(n_{part}/2)^{1/4}$ parameterization, as discussed in Sec.~\ref{partprod}. The $(n_{part}/2)^{1/4}$ parameterization is then elaborated to
\bea \label{npart}
\left(\frac{n_{part}}{2}\right)^{1/4} \hspace{-0.1in} &=& \frac{\sigma }{ \sigma_0} \left(\frac{n_{part,p}}{2}\right)^{1/4} + \left(1-\frac{\sigma }{ \sigma_0}\right)  \left(\frac{n_{part,0}}{2}\right)^{1/4} \nonumber \\
&+& \frac{0.05}{2.6}  \cos\left(2.6\left[(n'_{part}/2)^{1/4}-2.3\right]\right) \\ \nonumber
 &-& \frac{0.05}{2.6} \cos\left(2.6\left[(n_{part,p}/2)^{1/4}-2.3\right]\right),
\eea
where $(n'_{part}/2)^{1/4}$ is defined by the linear parameterization (first line), $n_{part,0}/2 = 191$ at $\sqrt{s} =$ 200 GeV and $n_{part,p}/2 = 0.45$. 

The solid and dash-dot curves in Fig.~\ref{fig5b} agree well, deviating from the lower dotted reference line (endpoint at $n_{part}/2 = 1/2$) by a small curvature concave downward due to the sinusoid component.  In the peripheral region (see inset) the thin solid line (parameterization without fluctuations) agrees almost exactly with the thick dash-dot line and solid dots from method 2) of the power-law integration, which fully reflects the participant-scaling hypothesis. The full parameterization  (thicker solid line in inset) includes an accommodation for $n_{ch}$ fluctuations [Eq.~(\ref{fluct})] such that $n_{part}/2 \rightarrow 1$ for peripheral collisions ({\em cf.} App.~\ref{running}).

The solid and dash-dot curves in Fig.~\ref{fig5c} also agree well. The definition of running integration is simpler for $n_{bin}$, since the middle Riemann sum on $n_{bin}$ is consistent with binary-collision scaling and bin averaging. Because the $d\sigma/d n_{bin}^{1/6}$ distribution in Fig.~\ref{fig5a} (right panel) begins well below the mean value and has a significant positive slope, the corresponding running integral in  Fig.~\ref{fig5c} (dash-dot curve) has a significant curvature concave downward which we accommodate by a modification of the linear power-law reference,
\bea \label{nbin}
n_{bin}^{1/6} =   \left(\sigma / \sigma_0 \right)^{0.965} \cdot n_{bin,p}^{1/6} + (1-\sigma / \sigma_0)^{1.035} \cdot n_{bin,0}^{1/6}.
\eea
The exponents on the cross-section factors add the necessary curvature to the parameterization, as shown by the close agreement between solid and dash-dot curves.

The final parameterizations in Figs.~\ref{fig5b} and \ref{fig5c} agree to $\sim 1$\% with the power-law integrals of the Glauber Monte Carlo data, except for the peripheral region where effects of multiplicity fluctuations are modeled in the full parameterization. To obtain the asymptotic approach to unity required for $n_{part}/2$ and $n_{bin}$ in peripheral A-A and N-N collisions we use
\bea \label{fluct}
x \rightarrow (1+x^a)^{1/a}
\eea
with $a = 8$ for $x = (n_{part}/2)^{1/4}$ and $a = 15$ for $x = n_{bin}^{1/6}$. The choices for $a$ depend on observed N-N multiplicity fluctuations as discussed in App.~\ref{running}. 

Also plotted in Figs.~\ref{fig5b} and \ref{fig5c} are results from an independent Monte Carlo Glauber analysis (nine solid points) based on bin averages on $n_{part}$ and $n_{bin}$~\cite{starglaub}. For $(n_{part}/2)^{1/4}$ in Fig.~\ref{fig5b}  the results agree with the power-law parameterizations for the most central collisions but significantly disagree for more-peripheral collisions. The points are consistent with the linear parameterization of $(n_{part}/2)^{1/4}$ with endpoint $n_{part,p}/2 = 3/4$ (upper dotted line). The dashed curves which pas through the solid points represent alternative running-integral definitions on $n_{part}$ described above, with endpoints at $(3/4)^{1/4}$. We conclude that the incorrect endpoint is a consequence of bin averaging on $n_{part}$ rather than $n_{part}/2$.

The $n_{bin}^{1/6}$ data (solid points) from~\cite{starglaub} in Fig.~\ref{fig5c}  seem to be consistent with endpoint $n_{bin,p} = 3/4$ (upper dotted line). However, although the points are slightly higher than the solid and dash-dot curves, both data and curves are displaced from the lower dotted line with endpoint 1/2 by a curvature resulting from the structure of the differential cross section of Fig.~\ref{fig5a} (right panel). Thus, the $n_{bin}$ data points are probably consistent with the correct endpoint 1/2.

 \subsection{Power-law mean path length $\nu$}

We have obtained precise parameterizations for $(n_{part}/2)^{1/4}$ and $n_{bin}^{1/6}$ {\em vs} fractional cross section. We now define participant path-length estimator $\nu$. The path-length concept originated with h-A experiments~\cite{busza}, for which $\nu$ was defined in terms of a hadron interaction length in nucleus A depending on the hadron-nucleon cross section and center-of-mass energy. The corresponding definition based on Monte Carlo Glauber parameters is $\nu = 2 n_{bin} / n_{part}$. Using the full power-law parameterizations from Eqs.~(\ref{npart}) and (\ref{nbin}) we have
\bea \label{nu}
\nu(\sigma/\sigma_0) \equiv \frac{2n_{bin}}{n_{part}} = \frac{\left(n_{bin}^{1/6}\right)^6}{\left([n_{part}/2]^{1/4}\right)^4}.
\eea
In Figs.~\ref{fig5b} and \ref{fig5c}  the $n_{part}/2$ and $n_{bin}$ parameterizations (solid curves) which accommodate $n_{ch}$ fluctuations are constrained by Eq.~(\ref{fluct}) so that  $n_{part}/2 $ and $ n_{bin} \rightarrow 1$ as $1 - \sigma/\sigma_0 \rightarrow 0$. Those transitions must be coordinated so that $\nu \rightarrow 1$ smoothly as well. 

\begin{figure}[h]
\includegraphics[width=3.3in,height=1.7in]{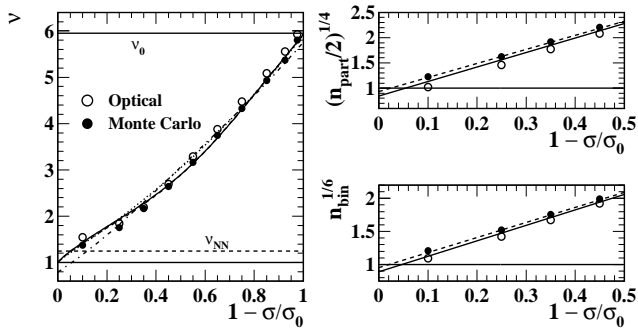}
\caption{Left panel: Participant mean path length $\nu$ {\em vs} fractional cross section. The solid curve is derived from the parameterizations of the running integrals in Figs.~\ref{fig5b} and \ref{fig5c} (solid curves), and the dotted curve is the same but without the $n_{part}$ sinusoid. The dash-dot curve is approximation $(n_{part}/2)^{1/3}$. The points represent two binned Glauber results~\cite{starglaub}. 
Right panels: Detailed views of Glauber power-law trends for peripheral collisions. The solid lines  with endpoints at $(1/2)^{1/4}$ and $(1/2)^{1/6}$ are derived from the dash-dot running integrals in Figs.~\ref{fig5b} and \ref{fig5c}. The points are from optical (open) and Monte Carlo (solid) Glauber simulations~\cite{starglaub}. The dashed lines extrapolate the solid points to endpoints at $(3/4)^{1/4}$ and $(3/4)^{1/6}$.  
\label{fig6}}
\end{figure}

Fig.~\ref{fig6} (left panel) shows $\nu(\sigma/\sigma_0)$ from Eq.~(\ref{nu}) as the solid curve. Maximum value $\nu_0 \sim 6$ corresponds to $b = 0$ and $\sigma_{NN}$ for 200 GeV N-N collisions ($n_{bin} \propto \sigma_{NN}$ for central collisions). $\nu_{NN} \sim 1.25$ corresponds to the centrality ($\sim$ 0.06) at which $n_{part}/2 \sim n_{bin} \sim 1$ ({\em cf.} solid curves, right panels). The dotted curve (barely visible near the center) shows the result when the sinusoid is omitted from the $n_{part}/2$ parameterization. The dash-dot curve $(n_{part}/2)^{1/3}$, a fair approximation to $\nu$ except in the most peripheral region, is used to approximate $\nu$ in Eq.~(\ref{runint}). The dashed curve which underlies the solid curve results from changing both lower endpoint values for Eq.~(\ref{nu}) from 1/2 to 3/4. The changes produce no significant change in $\nu$, showing that $\nu$ provides strong {\em common-mode reduction} of sensitivity to binning and averaging schemes. 

The extrapolation endpoints of the linear $n_{part}/2$ and $n_{bin}$ power-law trends are both $\sim 1/2$. However, the physical limit for each parameter is 1. Resolution of that conflict is addressed in App.~\ref{running}. It implies however that $\nu = 1$ does not represent the point on $\nu$ corresponding to a single N-N collision. To locate that point we make the following argument. The fractional cross section corresponding to $n_{part}/2 = 1$ in the linear power-law case is given by
\bea
\left\{1 - \frac{\sigma }{ \sigma_0}\right\} _{NN}  \hspace{-.1in} &=& \frac{1 - (n_{part,p}/2)^{1/4}}{(n_{part,0}/2)^{1/4} - (n_{part,p}/2)^{1/4}}.
\eea
With $n_{part,p}/2 \sim 1/2$ and $n_{part,0}/2 = 191$ we obtain $\left\{1 - \frac{\sigma }{ \sigma_0}\right\} _{NN} \sim 0.06$. $\nu_{NN} \sim 1.25$ is therefore the location on $\nu$ of the centroid of the single N-N multiplicity distribution, as noted in Fig.~\ref{fig6} and subsequent plots.

Participant mean path length $\nu$ is the ideal centrality measure for A-A collisions, providing precise visual tests of N-N linear superposition relative to a two-component combination of participant and binary-collision scaling. The per-participant yield of $n_{ch}$, $p_t$ or $E_t$ plotted {\em vs} $\nu$, for a simple combination of participant and binary-collision scaling, should exhibit a linear increase with $\nu$ relative to a constant background~\cite{nardi} ({\em cf.} App.~\ref{partprod2}). Deviations from linearity would reveal A-A medium effects, of central importance to RHIC physics. 

In contrast, the conventional centrality measure $n_{part} \sim 2\nu^3$ is a very nonlinear centrality measure, and binary-collision trends emerge as $n_{part}^{1/3}$ curves. As illustrated in Sec.~\ref{glauberr} the nonlinearity of $n_{part}$ relative to the fractional cross section  compresses much of the cross section into a small region near zero. The lower 50\% of the cross section occupies less than 15\% of the $n_{part}$ range. Central collisions dominate the $n_{part}$ plotting format. Thus, the opportunity to discern subtle but important deviations from binary-collision scaling starting from N-N collisions (e.g., jet quenching and other medium effects) is abandoned.

\subsection{Comparison of optical, Monte Carlo and power-law Glaubers}
 
Fig.~\ref{fig6} (left panel) also compares optical and Monte Carlo Glaubers with the power-law parameterization for $\nu$ in Eq.~(\ref{nu}). The open circles and solid points are bin mean values from optical and Monte Carlo Glauber simulations respectively, presented in Tables II and III of~\cite{starglaub}.  The agreement is notable in view of the significant discrepancies between Glauber implementations in $n_{part}$ and $n_{bin}$ in the right panels. The solid curve from the power-law parameterization also agrees well with $\nu =  2\langle n_{bin}^{1/6} \rangle^6 /\langle n_{part}^{1/4} \rangle^4 $, since those quantities are nearly linearly related to the fractional cross section. As noted above, $\nu$ provides excellent {common-mode reduction} of systematic errors.

In Fig.~\ref{fig6} (right panels) the lower half of the centrality range is shown to increase sensitivity in the critical peripheral region. The optical Glauber quantities go to zero asymptotically by definition. The solid curves are the linear parameterizations (without fluctuations) derived from the running integrals of minimum-bias distributions in Fig.~\ref{fig5b} and \ref{fig5c}, with endpoint $(1/2)^{1/4}$. The dashed curves which agree with the solid points from~\cite{starglaub} have  endpoints $(3/4)^{1/4}$. The endpoint difference and the preferred endpoint choice are discussed in App.~\ref{running}. The difference between dashed and solid curves propagates to a $\sim 50$\% error in $n_{part}/2$ for peripheral collisions.

\section{Centrality determination}

We now return to the problem of relating measured data to A-A collision geometry. The relation between measured $n_{ch}$ and fractional cross section $\sigma / \sigma_{0}$ is 
defined by running integration of the minimum-bias data distribution. The multiplicity is in turn related to the Glauber geometry parameters through the fractional cross section. The power-law format, running integrals and extrapolation constraints for N-N collisions greatly reduce systematic uncertainties in collision geometry, especially for peripheral collisions.

\subsection{Conventional Centrality}

In a conventional centrality determination the raw minimum-bias distribution $dN_{evt} / dn_{ch}$ is corrected for event-trigger inefficiency (typically a small effect for all $n_{ch}$), vertex-reconstruction inefficiency (possibly large for peripheral collisions and small $n_{ch}$), tracking (particle-detection) inefficiencies and backgrounds ({\em e.g.,} beam-gas collisions and photo-nuclear excitations). Tracking inefficiencies depending on $n_{ch}$ distort the minimum-bias distribution (e.g., change the average slope of the power-law format from negative to positive), and may produce systematic errors in the centrality determination. The corrected and normalized distribution $d\sigma(n_{ch})/dn_{ch}$ or $d\sigma(n_{ch})/dn_{ch}^{1/4}$ should integrate to the total cross section $\sigma_{0}$ corresponding to the trigger definition.

As an example, we consider an analysis of 130 GeV RHIC data~\cite{starglaub}. We work backward from centrality bin-edge definitions to the running integral to the differential power-law distribution. Ten centrality classes were defined on total multiplicity $n_{ch}$ = $n_{h^+}$ + $n_{h^-}$ detected in one unit of pseudorapidity. The raw minimum-bias distribution $dN_{evt}/dn_{ch}$ was corrected for significant trigger and vertex-finding inefficiencies below $n_{ch} = 50$ as follows. The raw data were scaled to agree with Hijing in the trusted $n_{ch}$ interval [50,100] (compare to the [5,25] interval in~\cite{spectra1}) where the collision dynamics were said to be dominated by A-A geometry and well-described by the Hijing model. 

The trigger/vertex efficiency below $n_{ch} = 50$ was determined by the ratio of data to Hijing minimum-bias distributions, and the data were corrected by that ratio, giving a reported overall trigger/vertex efficiency of 94\%, with a 60\% efficiency below $n_{ch} = 5$. The corrected distribution was normalized to the 6.9 barns total cross section $\sigma_0$ at 130 GeV estimated from Hijing and partitioned into ten bins according to the integrated fractional cross section. The total of beam-gas and photonuclear contributions to the most-peripheral 20\% bin was estimated to be 30\% according to~\cite{starglaub}, and that bin was therefore excluded from further analysis.

\begin{figure}[h]
\includegraphics[width=3.3in,height=1.65in]{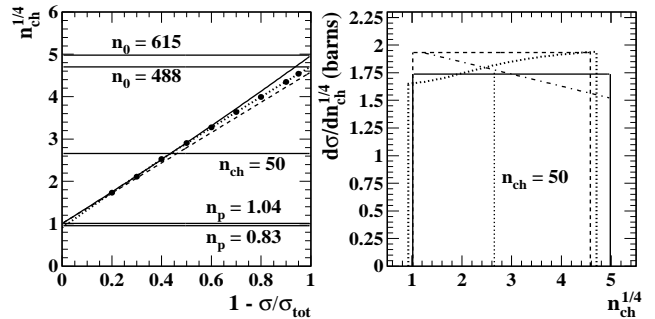}
\caption{Left panel: The points are bin edges from a conventional centrality definition for uncorrected particle multiplicities $n_{ch}$~\cite{starglaub}. The dotted curve is a parameterization of the points. The dashed curve is the participant-scaling reference for corrected $n_{ch}$.  The solid line (power-law reference) is an estimate of the corrected data trend for 130 GeV. Right panel: The corresponding power-law form of the differential cross section inferred from data (dotted curve and lines), the participant-scaling reference (dashed rectangle) and the estimated power-law reference for corrected data (solid rectangle). 
\label{fig7}}
\end{figure}

In Fig.~\ref{fig7} (left panel) the points represent reported centrality bin edges on {\em uncorrected} $n_{ch}$ in $|\eta| < 0.5$ and $p_t > 0.1$ GeV/c from Table I of~\cite{starglaub}. We infer upper endpoint $n_0 = 488$ for uncorrected data by extrapolating the bin-edge positions with a parameterization (dotted curve). We estimate the lower endpoint as $n_p = n_{part,p} / 2\cdot n_{NN} = 0.45\times 2.3 = 1.04$ for corrected $n_{ch}$ based on the expected 130 GeV p-p yield in a pseudorapidity acceptance of one unit~\cite{fabe}. The TPC tracking efficiency was reported to be 80\%, corresponding to $n_p = 0.83$ for uncorrected data.  The dashed line represents the participant-scaling reference for corrected data and the solid curve estimates the power-law reference for corrected data.

In Fig.~\ref{fig7} (right panel) the dotted lines and curve reconstruct the uncorrected differential cross section by differentiating the parameterization of the data points (dotted curve) in the left panel. The dashed rectangle represents the participant-scaling reference for corrected data, based on $n_{NN}  = 2.3$ and $\sigma_0 = 6.9$ barns, and the solid rectangle represents an estimate of the power-law reference for corrected data with $n_0 =1.4\,n_{part,0}/2 \cdot n_{NN}  = 615$. The dash-dot line estimates the expected trend of the corrected differential cross section and, given the 2$\times$ factor for $h^- \rightarrow n_{ch}$, compares well with Fig.~\ref{fig3} (right panel) for the $h^-$ yield from the same 130 GeV data~\cite{spectra1}.

The uncorrected distribution represented by the dotted lines and curve from~\cite{starglaub} differs markedly from the corrected results from~\cite{spectra1} represented by the dash-dot line. The reason is a strong $n_{ch}$ dependence of the tracking efficiency. The estimated 80\% TPC tracking efficiency is not uniform on $n_{ch}$. A TPC tracking efficiency typically decreases significantly with increasing $n_{ch}$ due to losses by cluster/track merging for larger track density. The result is the difference between uncorrected data (points and dotted curve) and estimated corrected data (solid line) in the left panel. The inefficiency variation {\em reverses the curvature} of the running integral in the left panel and changes the sign of the slope at mid-centrality in the right panel ({\em cf.} App.~\ref{partprod2}).

\subsection{Power-law Centrality}

\begin{figure}[h]
\includegraphics[width=1.65in,height=1.7in]{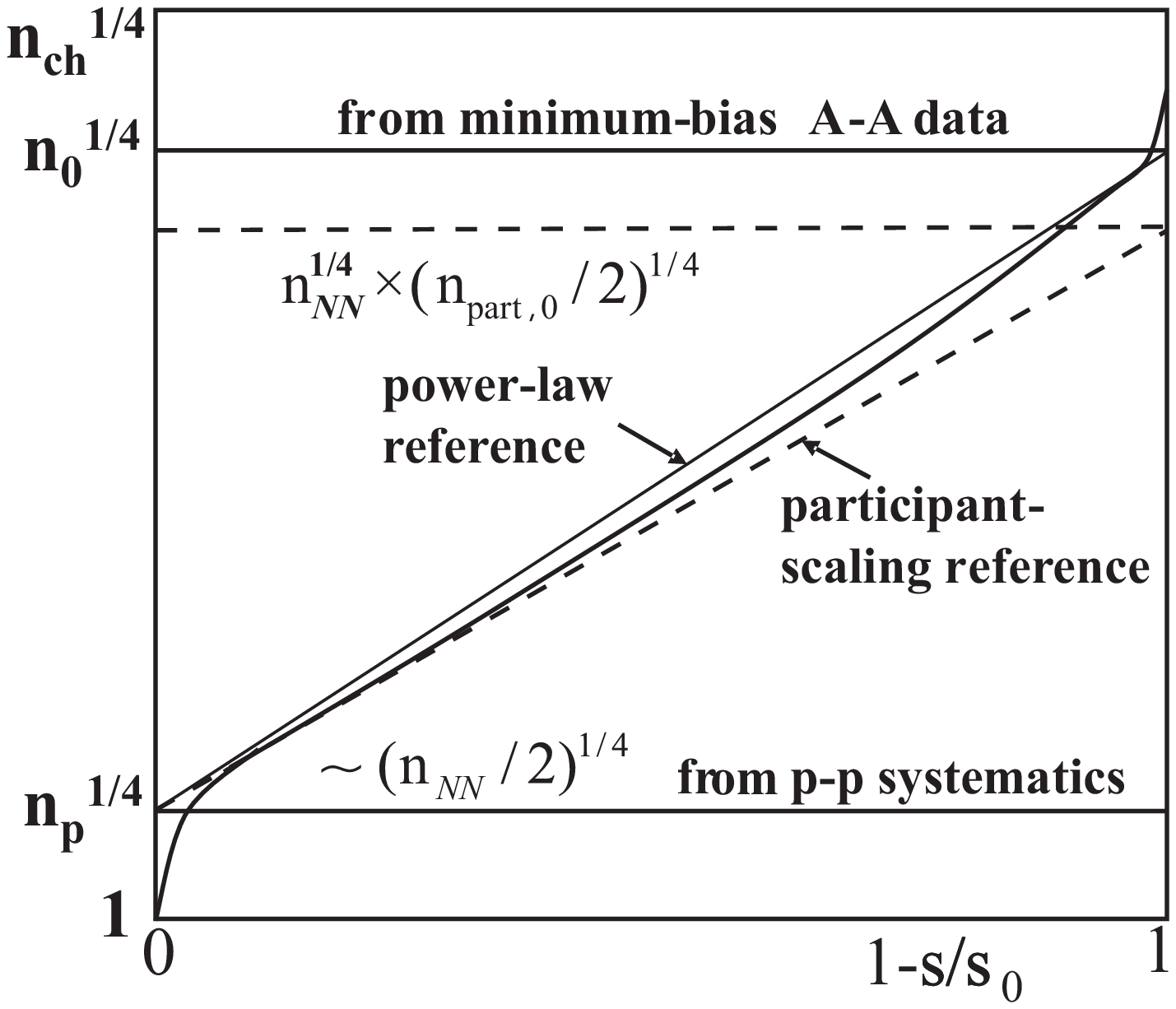}
\includegraphics[width=1.65in,height=1.7in]{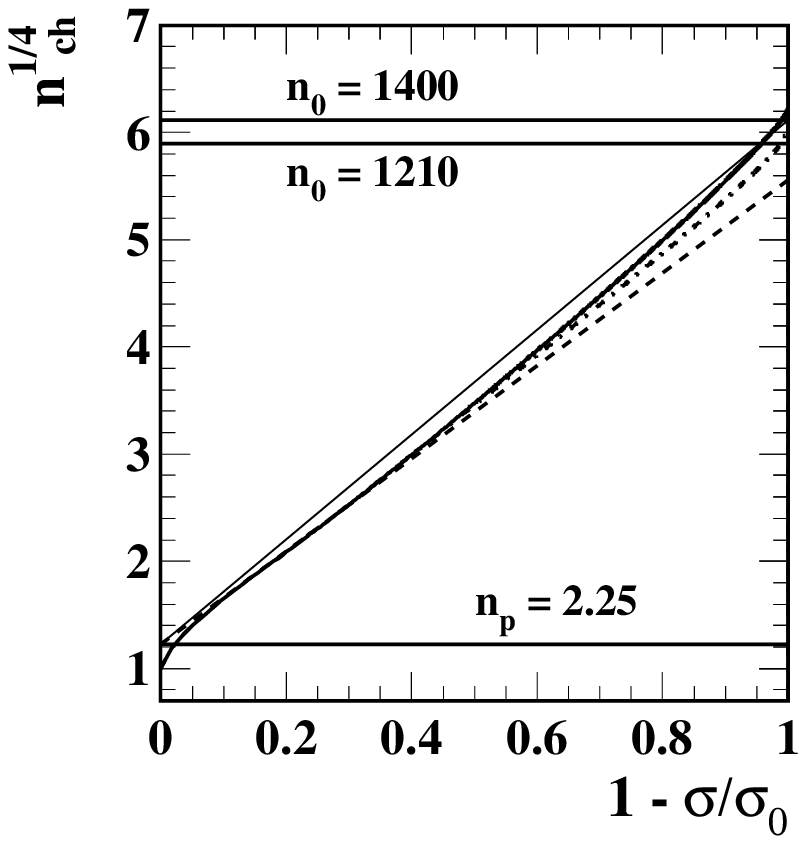}
\caption{Left panel: A cartoon of a running integral of the power-law differential cross section showing the principal features. Right panel: Running integrals of the differential cross sections in Fig.~\ref{fig2} (left panel) for quench-off (dash-dot) and quench-on (thicker solid) Hijing. The dashed line is the participant-scaling  reference. The thinner solid line is the power-law reference. 
\label{fig8}}
\end{figure}

Power-law centrality determination employs a running integral of the power-law differential cross-section distribution, removing the restriction to fixed centrality bins and a special $n_{ch}$ definition. The general features of the running integral are represented by the cartoon in Fig.~\ref{fig8} (left panel). The participant-scaling reference is the dashed line running from $n_p^{1/4} \sim (n_{part,p}/2 \cdot n_{NN})^{1/4}$ to $(n_0^*)^{1/4} \equiv  (n_{part,0}/2 \cdot n_{NN})^{1/4}$ on the vertical $n_{ch}^{1/4}$ scale and from 0 to 1 on fractional cross section $1 - \sigma/\sigma_0$. The power-law reference is the thin solid line running from $n_p^{1/4}$ to $n_0^{1/4}$. Those lines correspond to dashed and solid rectangles respectively in previous differential cross-section plots.

Endpoints $n_p$, $n_0$ and the slope at the midpoint of the differential cross section determine the shape of the running integral (solid curve). The differential endpoints determine the mean slope of the running integral. The differential slope determines the curvature of the running integral at its midpoint. The slopes at the ends of the differential distribution (from fluctuations) determine the curved segments at the ends of the running integral. The fluctuation shape near $n_p$ also depends on the skewness of the N-N multiplicity distribution, which causes a shift of $n_p$ slightly below $n_{NN} / 2$. 

Fig.~\ref{fig8} (right panel) shows running integrals of the Hijing quench-on (solid) and quench-off (dash-dot) differential power-law distributions in Fig.~\ref{fig2} (left panel). Running integration substantially reduces statistical noise~\cite{ppprl}. Dashed line $n_{NN}^{1/4} \cdot (n_{part}/2)^{1/4}$ represents participant scaling (particle production proportional to participant-pair number). Deviations from the power-law reference (thin solid line) should consist mainly of the small curvature in the central region corresponding to the slope of the differential distribution. Excursions near the endpoints correspond to multiplicity fluctuations. 

The curvatures for the Hijing data are concave upward, corresponding to the negative slope of the differential distributions in Fig.~\ref{fig2} (left panel). Given that $n_p \sim n_{NN}/2$ (exact for a symmetric N-N multiplicity distribution), knowledge of the NSD p-p multiplicity to $\sim$10\% means that the centrality accuracy for peripheral A-A collisions is $\sim 1$\%. For Hijing, the relation is $n_p \sim 0.45\, n_{NN} = 2.25$ due to the skewness of the N-N NBD distribution ({\em cf.} App.~\ref{running}). In Fig.~\ref{fig8} (right panel) $n_p = 0.45 \cdot 5 = 2.25$ results in excellent agreement between Hijing data (curves) and the participant-scaling reference (dashed line) for peripheral collisions.

\section{Particle, $p_t$ and $E_t$ production}

One purpose of A-A centrality determination is to establish how various quantities are `produced' in the final state from a multitude of initial N-N (or parton-parton) collisions, and how (or if) the production mechanisms change with A-A collision geometry. Production of final-state hadrons, transverse momentum $p_t$ and transverse energy $E_t$ is determined by initial-state parton scattering, parton dissipation in the bulk medium, the dynamics of the bulk medium and the hadronization process (scattered-parton and bulk-medium fragmentation). Change can be defined relative to a linear superposition hypothesis: a linear combination of products from the number of N-N collisions predicted by combining the Glauber and two-component models. 

$n_{ch}$, $p_t$ and $E_t$ per participant pair should exhibit a combination of participant scaling (independent of $\nu$) and binary-collision scaling (proportional to $\nu$), plus medium effects which may be substantial. The differential procedure consists of two parts: 1) determine the ratio of a produced quantity to the number of N-N participant pairs, and 2) plot the variation of that ratio {\em vs} path length $\nu$, the number of N-N collisions per participant pair. That procedure can be illustrated with the Hijing Monte Carlo and RHIC data. In contrast to particle number production, the $p_t$ and $E_t$ minimum-bias distributions involve continuous variables which require some differences in integration technique. In the case of $E_t$ production we compare conventional and power-law methods applied to RHIC data.

\subsection{Particle production} \label{partprod}

The two-component model of nuclear collisions~\cite{nardi,ppprl} compares the fraction of particle production in A-A collisions due to participant scaling $n_{ch} \propto n_{part}/2$ (soft component) and the fraction due to binary-collision scaling $n_{ch} \propto n_{bin}$ (hard component). That decomposition may separate contributions from initial-state parton scattering and fragmentation, which should scale as the latter, from bulk-medium hadronization which may scale as the former. By comparing deviations from participant scaling with a linear binary-collision reference on participant path-length $\nu$, modifications to parton scattering and fragmentation by the QCD medium (e.g., jet quenching) may be analyzed.

Measurement of particle production in the form $2/n_{part}\cdot n_{ch}$ is a stringent test of centrality determination, since the relation of $n_{ch}$ to $n_{part}/2$ is sensitive to small relative errors in $\sigma / \sigma_{0}$ inferred separately in the data and Glauber contexts. Conventional centrality methods without extrapolation constraints entail uncertainties for peripheral collisions  large enough that particle-production studies with conventional methods have not been attempted for the 20\% most peripheral collisions. The power-law method opens that region to precise study.

In the simplest power-law method $n_{part}^{1/4}$ and $n_{bin}^{1/6}$ {\em vs} $\sigma / \sigma_{0}$  are represented by linear parameterizations sufficiently accurate for centrality determination. However, the sinusoid on $d\sigma/dn_{part}^{1/4}$ is $0.05\,\sin\{2.6(n^{1/4}_{part} - 2.3)\}$ (relative to the mean power-law cross section). The resulting relative deviation of $n_{part}^{1/4}$ {\em vs} $\sigma(n^{1/4}_{part}) / \sigma_{0}$ from a power-law trend is 0.014 at the midpoint, an error of 1.5\% in centrality determination. However, particle-production studies involve $n_{part}/2$ and not $(n_{part}/2)^{1/4}$. The sinusoid contributes a deviation in $n_{part}/2$ of 6\% from the power-law reference, sufficient to require including the sinusoid in particle-production studies. Therefore, the sinusoid in Fig.~\ref{fig5a} (left panel) should be included in the parameterization of $n_{part}^{1/4}$ {\em vs} $\sigma / \sigma_{0}$. 

\begin{figure}[h]
\includegraphics[width=1.65in,height=1.6in]{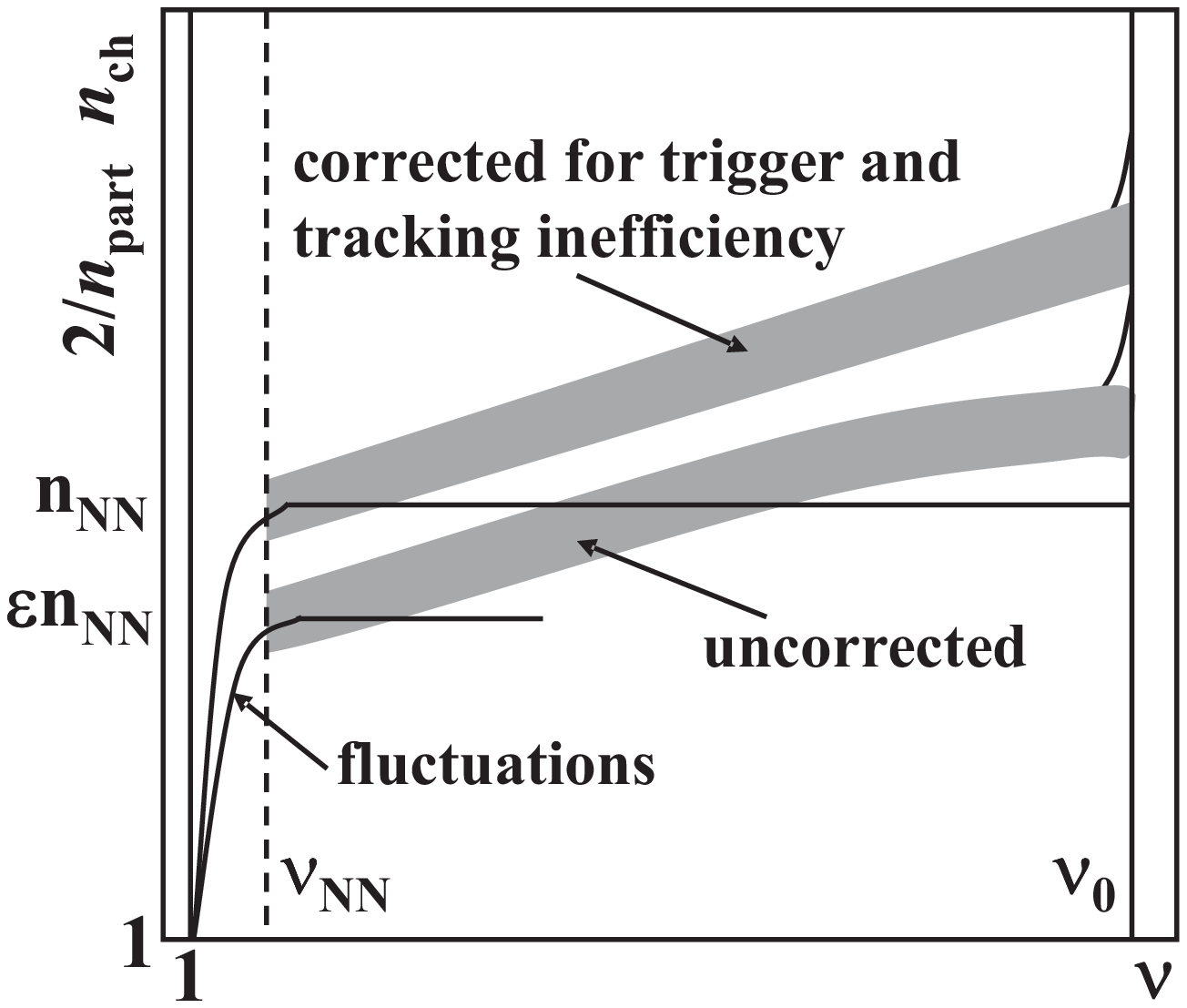}
\includegraphics[width=1.65in,height=1.625in]{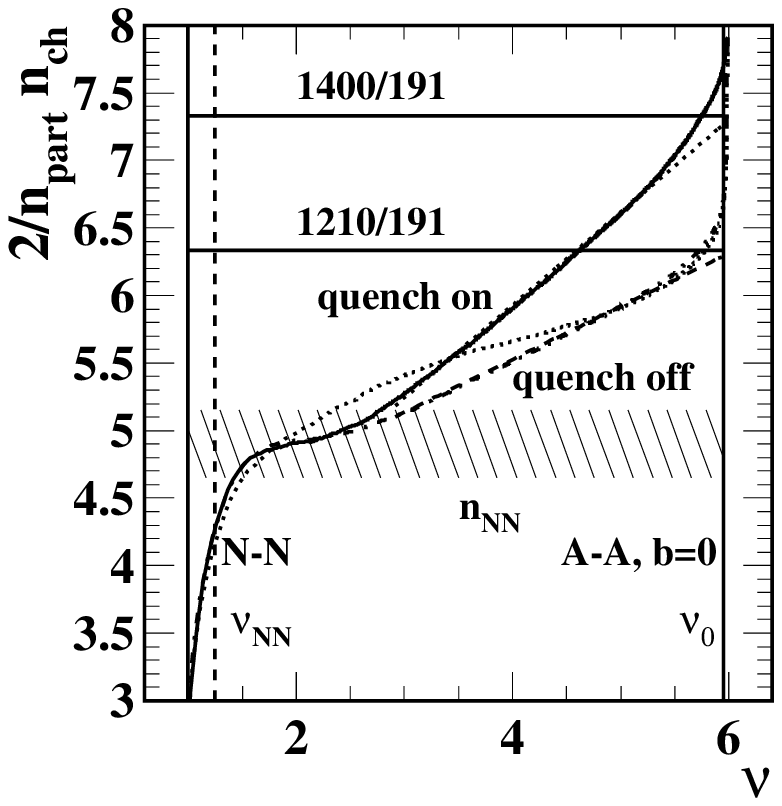}
\caption{Left panel: A cartoon of particle production per participant pair {\em vs} mean participant path-length $\nu$ for uncorrected (lower) and corrected (upper) data. The effects of multiplicity fluctuations are illustrated at the ends.  Right panel: Particle production curves for Hijing quench-off (dashed) and quench-on (solid) collisions. The dotted curve illustrates the consequence of omitting the sinusoid in the  $n_{part}/2$ parameterization (linear approximation).  The hatched region represents the N-N (NSD p-p) multiplicity in the acceptance at 200 GeV ($n_{NN} \sim 5$ for Pythia). The horizontal solid lines represent $2 n_0 / n_{part,0}$.  
\label{fig11}}
\end{figure}

Fig.~\ref{fig11} (left panel) is a cartoon of $2/n_{part}\cdot n_{ch}$ {\em vs} mean participant path length $\nu$ that combines $n_{ch}^{1/4}$ {\em vs}  $1 - \sigma / \sigma_{0}$ obtained from a data minimum-bias distribution with $n_{part}^{1/4}$ and $\nu$ {\em vs}  $1 - \sigma / \sigma_{0}$ obtained from a Glauber Monte Carlo. The lower band represents data uncorrected for tracking inefficiency (including $n_{ch}$ dependence which produces the curvature), and the upper band represents corrected data. $n_{NN}$ is the N-N ($\sim$NSD p-p) multiplicity in the acceptance derived from separate experiments, and $\epsilon$ is the tracking efficiency for peripheral collisions. The horizontal line extending to $\nu_{0}$ ($b=0$) is the participant-scaling reference. The curves represent deviations from power-law trends due to fluctuations.

Fig.~\ref{fig11} (right panel) shows the result of a particle-production study of the Hijing Monte Carlo, with $n_{part}(b)/2$ {\em vs} $\sigma(b)$ derived from the Glauber parameterization with sinusoid (solid curve) in Fig.~\ref{fig5b} and $n_{ch}$ {\em vs} $\sigma(n_{ch})$ derived from the solid and dash-dot curves in Fig.~\ref{fig8} (right panel). We plot $2/n_{part}\, \Delta \eta\, dn_{ch}/d\eta$, with $\Delta \eta = 2$. The curves correspond to quench-off (dash-dot) and quench-on (solid) Hijing. The dotted curve shows the result of omitting the sinusoid from the $n_{part}$ parameterization---a 6\% systematic error for particle production. 
Parameter $a$ in the Glauber parameterization which represents fluctuations affects the curves only near $\nu = 2$. For $1/a \rightarrow 0$ the slopes of the curves become negative near $\nu = 2$ because $n_{part}/2$ without fluctuations decreases too rapidly with decreasing $\nu$.  The adopted value $a=8$ for $n_{part}/2$ was the largest value for which the slopes of the curves were non-negative near $\nu = 2$. 

The hatched region (mean value and error band) estimates participant scaling from $dn_{ch}/d\eta \sim$ 2.5 for NSD p-p collisions ($\nu \approx 1.25$) at 200 GeV integrated over two units of pseudorapidity, consistent with SP\=PS p-\=p data~\cite{ua1}.  The width of the error band represents the systematic uncertainty propagated from the 10\% uncertainty in $n_{part,p}/2 = 0.45\pm0.05$ which produces a 5\% maximum uncertainty in $2/n_{part} \, dn_{ch}/d\eta$ confined to the region around $\nu = 2$. The curves above $\nu = 2.5$ are insensitive to that endpoint uncertainty.

Interpretation of the particle-production evolution from Fig.~\ref{fig11} (right panel)  is straightforward. Below $\nu \sim 2$ there is participant scaling and fluctuations. Above $\nu \sim 2$ the quench-off trend increases with path length, presumably reflecting increased hadron production from \mbox{(mini-)jets} produced in multiple N-N collisions. The quench-on trend shows additional increase above $\nu \sim 2.5$. Presumably, parton energy loss (quenching proportional to final-state path length, also $\propto \nu$) is converted to additional hadron production. Unprecedented access to such a detailed picture of particle production illustrates the importance of the power-law centrality method.

\subsection{Transverse momentum $p_t$ production}

\begin{figure}[h]
\includegraphics[width=1.7in,height=1.6in]{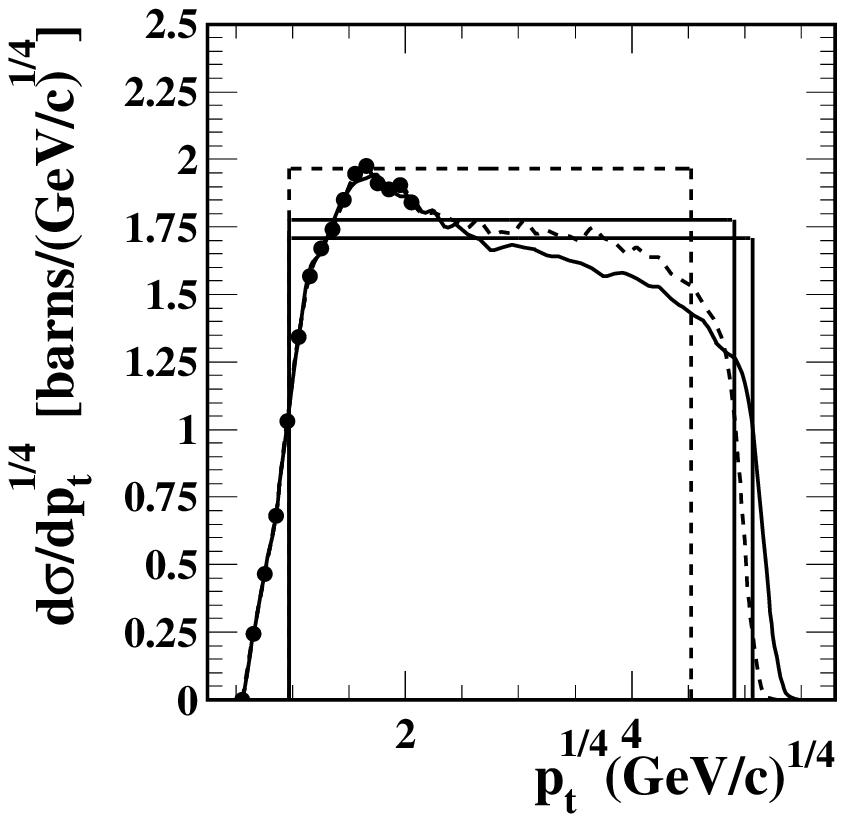}
\includegraphics[width=1.6in,height=1.6in]{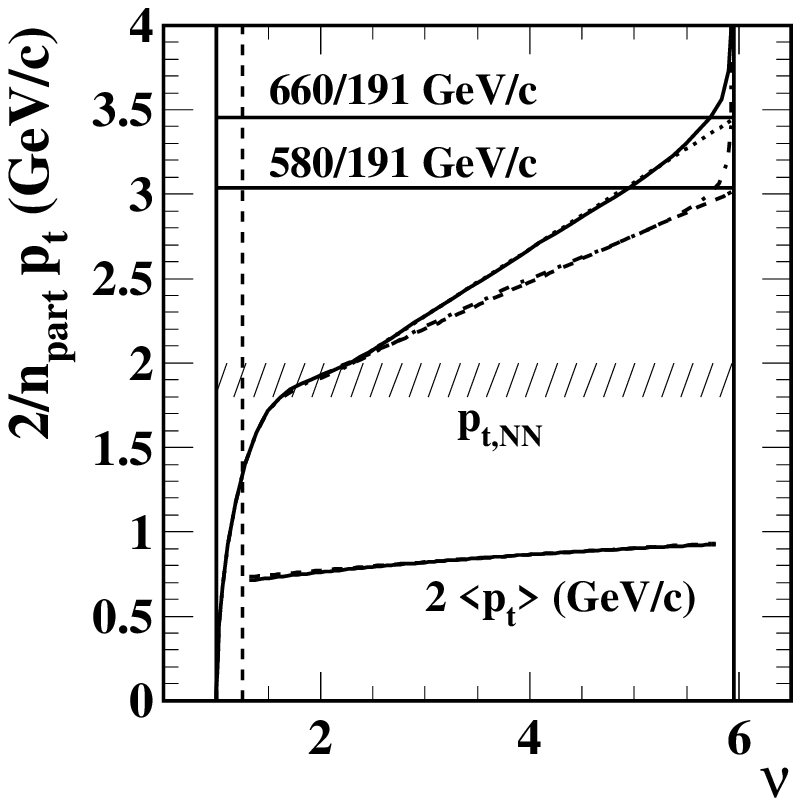}
\caption{Left panel: The power-law form of the minimum-bias distribution on total $p_t$ within $|\eta| < 1$ for quench-on (solid) and quench-off (dashed) Hijing. The points at the left end illustrate the binning scheme uniform on $p_t^{1/4}$.  Right panel: The total $p_t$ {\em per-participant pair} within the acceptance {\em vs} mean participant path length $\nu$ for the two Hijing classes. The lower solid and dash-dot curves represent twice the {\em per-final-state-particle} mean $p_t$ $ 2\,\langle p_t \rangle$ for the two cases.  
\label{fig12}}
\end{figure}

We next study the  power-law distribution and $p_t$ production. To obtain the distributions in Fig.~\ref{fig2} (left panel) the conventional minimum-bias distribution was first formed on integer multiplicity $n_{ch}$, then rebinned onto $n_{ch}^{1/4}$. For continuous variable $p_t$, in contrast, $p_t^{1/4}$ was binned into 50 equal bins, providing adequate resolution for end-point structure. Event numbers were accumulated directly into the bins on $p_t^{1/4}$ (not $p_t$). Fig.~\ref{fig12} (left panel) shows $d\sigma / dp_{t}^{1/4}$ {\em vs} $p_{t}^{1/4}$. The points on the left edge illustrate the uniform bin spacing and edge resolution. The upper half-maximum point $p_{t0}$ estimates the correspondence on total $p_{t}$ of $b=0$ and $n_{part,0}/2$, and the lower half-maximum point $p_{tp}$ estimates $n_{part,p}/2 \cdot p_{t,NN}$, roughly half the total $p_t$ in the acceptance for N-N collisions. The endpoints for Hijing are $p_{tp} \sim 0.86 \sim 0.45\, p_{tNN}$ GeV/c and $p_{t0} \sim 580~(660)$ GeV/c for quench-off (quench-on) events. 

In Fig.~\ref{fig12} (right panel) we plot the total {\em per-participant-pair} quantity 
\bea
2/n_{part}\, \Delta \eta \, dp_t /d\eta &\equiv& 2/n_{part}\, \Delta \eta\, dn_{ch}/d\eta \cdot \langle p_t \rangle
\eea 
within acceptance $\Delta \eta = 2$. The labeled horizontal solid lines represent $2 p_{t,0}/n_{part,0}$ for quench-on and quench-off events. The hatched region represents $p_{t,NN} \sim 1.9$ GeV/c $\sim n_{NN} \cdot 0.38$ GeV/c. The left endpoint $p_{t,p} \sim 0.86$ GeV/c in the left panel is about 0.45$\times$ the N-N value in the right panel, as expected. The slope for quench-on Hijing corresponds to $x_{p_t} \sim 0.16$, compared to $x_{n_{ch}} \sim 0.08$. That difference is qualitatively consistent with the trend of increasing $\langle p_t \rangle$ discussed below: $p_t$ increases with centrality faster than $n_{ch}$, and we expect $x_{p_t} > x_{n_{ch}}$.

The (overlapping) solid and dash-dot curves below the hatched region show the ratio of $p_t$ production in Fig.~\ref{fig12} and multiplicity $n_{ch}$ production in Fig.~\ref{fig11}. Quench-off and quench-on events exhibit the same $\langle p_t \rangle$ variation, roughly consistent with trends for RHIC data at 130 and 200 GeV~\cite{rhicdat} and 200 GeV p-\=p results ($\nu \sim 1.25$)~\cite{ua1}. 

Particle and $p_t$ production systematics from Hijing collisions are closely related to $\langle p_t \rangle$ {\em fluctuations} and related $p_t$ {\em correlations} from that model representing minijet structure~\cite{hijsca}. The variation of Hijing per-particle $\langle p_t \rangle$ fluctuations with centrality was found to be small~\cite{QT}, in disagreement with RHIC $\langle p_t \rangle$ fluctuation measurements reported in~\cite{ptprl}. The weak centrality dependence of Hijing minijet-related $p_t$ angular correlations~\cite{hijsca} is also very different from corresponding RHIC data~\cite{ptsca}. Since those analyses are based on per-particle fluctuation and correlation measures we conclude that quench-off Hijing scales minijet production and total particle production almost identically. Hijing quench-off is simply a linear superposition of Pythia N-N collisions whose number follows a combination of participant and binary-collision scaling. In contrast, fluctuation and correlation analysis of RHIC data reveals that Au-Au collisions exhibit strong deviations from such linear superposition.

\subsection{Transverse energy $E_t$ production}

\begin{figure}[h]
\includegraphics[width=3.3in,height=3.2in]{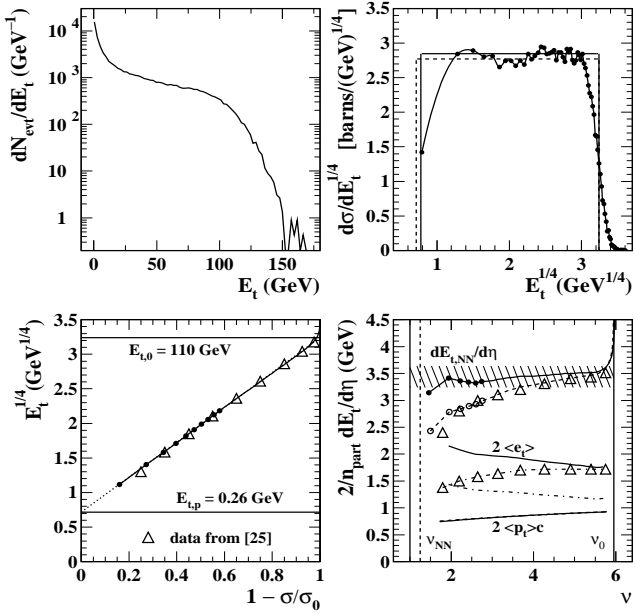}
\caption{Upper-left panel: 
The conventional semi-log form of the minimum-bias density distribution on total $E_t$ in an EMCal patch (see text).
Upper-right panel: 
The power-law form of the normalized minimum-bias density distribution on $E_t$ for data from~\cite{etpaper}.  
Lower-left panel:  Running integral of the distribution in the upper-right panel (solid curve and small points) compared to participant-scaling reference $ (n_{part}/2)^{1/4} \cdot E_{t,NN}$ (dotted line).
Lower-right panel: The density on $\eta$ of $E_t$ per participant pair. The solid and dashed curves are derived from the running integral in the lower-left panel with $n_{part}/2 $ parameterizations corresponding to intercepts $n_{part,p}/2 =$ 1/2 and 3/4 respectively.  The lower solid curves compare $E_t$ per charge hadron with $p_t$ per charged hadron. The open triangles in the lower panels are from~\cite{etpaper}. 
  \label{fig13}}
\end{figure}

Fig.~\ref{fig13} (upper-left panel) shows a semi-log minimum-bias distribution on transverse energy $E_{t}$ measured with an electromagnetic calorimeter (EMCal) patch~\cite{etpaper}. The data were corrected for trigger inefficiencies. The patch acceptance was 1/6 of 2$\pi$ azimuth over one unit of pseudorapidity [0,1]. The $E_t$ axis was uniformly binned (bin width 2.34 GeV) except for the lowest bin (width 1.56 GeV).
Bin sums were converted to densities $dN_{evt,i}/ dE_t$ for this analysis, dividing the $N_{evt,i}$ by bin widths $\delta E_{ti}$ to obtain the density distribution in the upper-left panel.

The $E_t$ minimum-bias distribution plotted in the power-law format in the upper-right panel was obtained by multiplying the upper-left distribution bin-wise by Jacobian $4 E_{ti}^{3/4}$ and normalizing to $\sigma_0 = 7$ barns (the default value assumed for this study). Because the binning is (mostly) uniform on $E_t$ the distribution suffers from sparse sampling at the low-$E_t$ end critical for centrality and $E_t$ production studies, and excessive sampling at the upper end. Uniform binning on $E_t^{1/4}$ is preferable, as illustrated by the $p_t$ analysis in Fig.~\ref{fig12} (left panel). 

The $E_t$ distribution is almost exactly power-law in form over the observed $E_t$ interval, consistent with participant scaling. The upper endpoint is $E_{to}^{1/4} = 3.24$  = (110 GeV)$^{1/4}$.  The lower endpoint must be estimated. The position of the first data point (roughly the half-maximum) at $E_{t}^{1/4} = 0.78$ (GeV)$^{1/4}$ establishes the rough estimate $E_{tp} = 0.37$ GeV. However, a lower endpoint derived from the observed upper endpoint and assuming pure participant scaling is $(0.45\cdot 110/191)^{1/4} = 0.26^{1/4} = 0.71$ (GeV)$^{1/4}$, with $E_{t,NN} = 0.58$ GeV also inferred. 

Given a common upper endpoint and using the two estimates the dashed rectangle in Fig.~\ref{fig13} (upper-right panel) is the participant-scaling reference, and the solid rectangle is the power-law reference. The lower endpoint should not be higher than that derived from the upper endpoint and a participant-scaling assumption. The power-law reference obtained from the lowest data point (solid rectangle) is obviously higher than the mean value of the data. Since the position of the single data point on the lower edge is not a precise estimate we adopt 0.26 GeV (and the dashed participant-scaling hypothesis) as the best model for the data distribution.

The lower-left panel shows the running integral of the minimum-bias data distribution in the upper-right panel. The half-maximum points inferred from the upper-right panel are represented by the horizontal lines. The diagonal dotted line $(E_{t,NN}\cdot n_{part}/2)^{1/4}$ with lower endpoint at $E_{tp}^{1/4} = (0.45\, E_{t,NN})^{1/4}$ is the participant-scaling reference. The solid curve and small points plotted at bin edges on $E_t^{1/4}$ represent the {recommended} running integral described in App.~\ref{runbin}. The excellent agreement with the participant-scaling reference (dotted line) is apparent. The open triangles are from~\cite{etpaper} and also agree well with the participant-scaling reference.

The lower-right panel shows $dE_t/d\eta$ per participant pair. As noted, a factor 6$\times$ was applied to the measured EMCal patch $E_t$ values to obtain $dE_t/d\eta$ values according to the definition of the patch acceptance. The solid curve and solid dots represent the recommended integration scheme and $N_{part} / 2$ parameterization from the present study. Those results are also consistent with participant scaling (hatched region), as expected from the uniformity of the power-law minimum-bias distribution. Assuming $E_{t,0} = E_{t,0}^*$, the constant value for all $\nu$ is $dE_t /d\eta = 6 \cdot E_{t,NN} = 3.46$ GeV.

The open triangles are data from~\cite{etpaper}. The systematic error bars in that paper (up to 25\% or $\sim 0.6$ GeV for the most peripheral points) are omitted in this comparison, since the same underlying minimum-bias data are used to compare different integration schemes and Glauber parameters. The points from~\cite{etpaper} (upper triangles) deviate significantly and systematically from the participant-scaling reference (hatched region) and the recommended running integral (upper solid curve). The dashed curve and open circles represent the correct $E_t^{1/4}$ running integral from the lower-left panel and the incorrect $N_{part/2}$ parameterization with lower endpoint at 3/4. That combination describes the points from~\cite{etpaper} well,  suggesting that the incorrect $n_{part}/2$ definition with lower endpoint at 3/4 contributed to the systematic error in~\cite{etpaper}. The systematic uncertainty represented by the hatched region is 5\%. whereas the uncorrected error in the triangle data points approaches 50\% at the N-N limit. This example illustrates the importance of the extrapolation constraints available within the power-law context.

$E_t$ and $p_t$ centrality trends for data and Hijing are said to agree in~\cite{etpaper}. However, comparing the {consistent} analyses in Fig.~\ref{fig12} (right panel) and Fig.~\ref{fig13} (lower-right panel) we find that the charged-particle $p_t$ and total $E_t$ (including neutrals) production trends are very different. The total $E_t$ (hadronic plus electromagnetic transverse energy) per participant pair is independent of centrality (indicative of participant scaling), whereas the number of charged hadrons per participant pair increases by 40\% from peripheral to central collisions at 200 GeV. Therefore, the total $E_t$ per charged hadron $\langle e_t \rangle$ defined in~\cite{etpaper} must {\em decrease}, whereas the conclusion in~\cite{etpaper} is that $\langle e_t \rangle$ increases substantially, similar to Hijing $n_{ch}$ and $p_t$ production. Solid curves for $\langle e_t \rangle$ (RHIC data) and $\langle p_t \rangle c$ (Hijing) are compared in the lower part of the bottom-right panel. The multiplicity trend we used to obtain consistency with~\cite{etpaper} at $\nu \sim 6$ is $dn_{ch} / d\eta = 2.95\, (1 + 0.08(\nu - 1))$, whereas we expect the prefactor to be 2.5 at 200 GeV. It is notable that the solid curves from the present study tend to converge. 

That $\langle e_t \rangle$-$\langle p_t \rangle c$ comparison is still not completely appropriate because the measured $E_t$ represents all particles including neutrals, whereas $\langle p_t \rangle$ represents only charge particles.  The $\langle e_t \rangle$ defined in~\cite{etpaper} in terms of $n_{ch}$ should instead be defined as $\langle e_t \rangle \equiv dE_t/d\eta \, \mbox{\large \bf / } \!\! dn_{tot}/d\eta$. If we adopt a factor 2/3 correction from $n_{ch}$ to $n_{tot}$, assuming hadrons are dominated by pions, we obtain the dash-dot curve in the lower-right panel of Fig.~\ref{fig13}, and the convergence of $\langle e_t \rangle$ and $\langle p_t \rangle c$ with increasing centrality toward 0.5 GeV is more apparent. The convergence is somewhat artificial (it could be better or worse) because Hijing does not necessarily model $\langle p_t \rangle$ from data correctly.
 
In App.~\ref{partprod2} we show algebraically that the four panels of Fig.~\ref{fig13} are redundant. The main experimental result of the analysis in~\cite{etpaper} is that $E_t$ production in Au-Au collisions at 200 GeV and mid-rapidity follows participant scaling ($x _{E_t}\sim 0$) within statistical errors. Some of the conclusions reported in~\cite{etpaper} are in error because the Glauber $n_{part}/2$ trend from~\cite{starglaub} used to calculate $2/n_{part}\, dE_t/d\eta$ is incorrect. The discrepancy is apparent in the right panels of Fig.~\ref{fig13}. The correct $n_{part}/2$ parameterization from the present analysis restores consistency, and the physics conclusions change significantly.

\section{Centrality Errors}

We summarize the systematic error sources for two centrality methods. We distinguish error sources for the Glauber model and for data, and for the conventional and power-law methods. The main technical issue is the correct relation of data and Glauber parameters to the {fractional} cross section $\sigma / \sigma_0$. Uncertainties in the {\em relative} centralities of Glauber model and data, most relevant for particle and $p_t/E_t$ production, may be significantly reduced by use of extrapolation constraints.

\subsection{Total cross section and error}

The total cross section for Au-Au collisions is estimated as follows. The Woods-Saxon matter distribution for Au has $r_0 = 6.5$ fm (including an estimated 0.1 fm from the neutron skin) and diffuseness $a = 0.5$ fm~\cite{miller}, implying a nominal nuclear edge at $r_0 + 2a = 7.5$ fm. The nucleus-nucleus cross section is then $\pi (2\cdot 7.5 \text{ fm})^2$/100 = 7.05 barns. An uncertainty of 0.1 fm in the edge radius results in an uncertainty of 0.2 barns in the total cross section. That $\pm 3$\% range encompasses all Glauber estimates for RHIC Au-Au collisions at 130 and 200 GeV. For  comparisons with published data we use stated total cross sections. For internal comparisons and illustration we adopt the default value $\sigma_0 = 7$ barns.

\subsection{Glauber-model errors} \label{glauberr}

We relate Glauber geometry parameters to data through their fractional cross-section dependencies. The error propagation from fractional cross section to Glauber parameters can be estimated from the linear power-law parameterizations. For $n_{part}$, $\delta n_{part} / n_{part} = [4(3.72 - 1)]\, \delta \sigma / \sigma_0$, and the coefficient in square brackets is 10.9. For $n_{bin}$, $\delta n_{bin} / n_{bin} = [6(3.23 - 1)]\, \delta \sigma / \sigma_0$, and the coefficient in brackets is 13.4. Both coefficients are large, implying the need for control of the fractional cross-section error at the 1\% level for effective geometry determination.  In the conventional approach that level of precision has not been achieved. 

\begin{figure}[h]
\includegraphics[width=3.3in,height=1.65in]{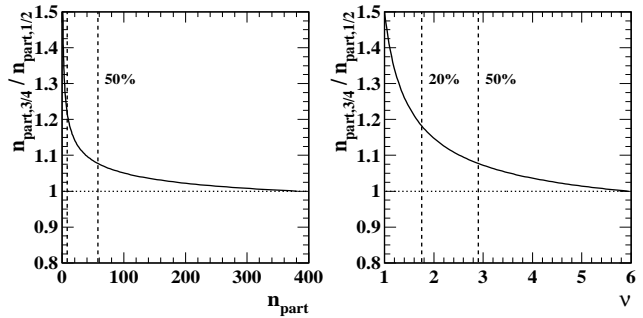}
\caption{The fractional systematic error in $n_{part}$ resulting from an incorrect lower endpoint (3/4 {\em vs} 1/2) plotted against $n_{part}$ (left panel) and $\nu$ (right panel), with the same two values of the fractional cross section indicated in each panel.
\label{fig16}}
\end{figure}

Absence of extrapolation constraints can result in large systematic errors, as sketched in Fig.~\ref{fig16}. The fractional error in $n_{part}$ resulting from an incorrect endpoint value increases to  50\% for peripheral collisions. In the conventional plotting format of the left panel ({\em cf.} Fig. 8 of~\cite{etpaper}) the impact of the systematic error (not uncertainty) is minimized by the nonlinear relation of $n_{part}$ to fractional cross section. The 20\% most peripheral collisions are omitted (below the left dashed line), and large systematic error bars (uncertainty estimates) are applied below 50\%. That strategy abandons all information on the transition from N-N to mid-peripheral A-A: how heavy ion collisions become distinct from elementary collisions. In contrast, the plotting format in the right panel provides precise access to the peripheral transition region, provided systematic errors are brought under control.

In the power-law context adequate control is accomplished by invoking {\em extrapolation constraints}:  aside from fluctuation effects both Glauber running integrals should extrapolate to endpoints $\sim$1/2 to be consistent with the power-law distributions on $n_{ch}$, $p_t$ and $E_t$. On the other hand, both Glauber parameters should extrapolate to 1 in the limit of N-N collisions or $\sigma/\sigma_{0} = 1$. Parameterizations satisfying those constraints reduce systematic uncertainties from the Glauber parameters to about 5\% in the troublesome peripheral region ($\nu \sim 2$).

As noted, systematic errors for path-length $\nu$ are much reduced because of common-mode error reduction. The relative error is $\delta \nu / \nu = \{6\cdot 2.23 / n_{bin}^{1/6} - 4\cdot 2.72 / (n_{part}/2)^{1/4} \}\, \delta \sigma / \sigma_0$. The coefficient in curly brackets has limiting values 2.5 (peripheral) and 1.25 (central). Thus, the error in $\nu$ relative to the fractional cross section is $O(1$-$2)$ compared to relative errors for  $n_{part}/2$ and $n_{bin}$ which are $O(10$-$15)$. With extrapolation constraints the systematic uncertainty in $\nu$ can typically be reduced to $< 2$\%, even in the peripheral region.

\subsection{Data errors}

Trigger and vertex-reconstruction inefficiencies, the latter typically significant for peripheral collisions, result in distortion of the minimum-bias distribution leading to systematic errors in the inferred centrality. The conventional method does not utilize critical {\em a priori} information available from p-p collisions which constrains the form of the minimum-bias distribution in the peripheral region. Without such constraints the fractional cross-section uncertainty can be 5-10\% for peripheral collisions, making the $n_{part}$ and $n_{bin}$ parameters meaningless in that region (50-100\% error). For that reason the 20\% most-peripheral part of the cross-section, which contains critical information on the transition from N-N to A-A collisions, is typically abandoned in the conventional approach. 

\begin{figure}[h]
\includegraphics[width=1.65in,height=1.5in]{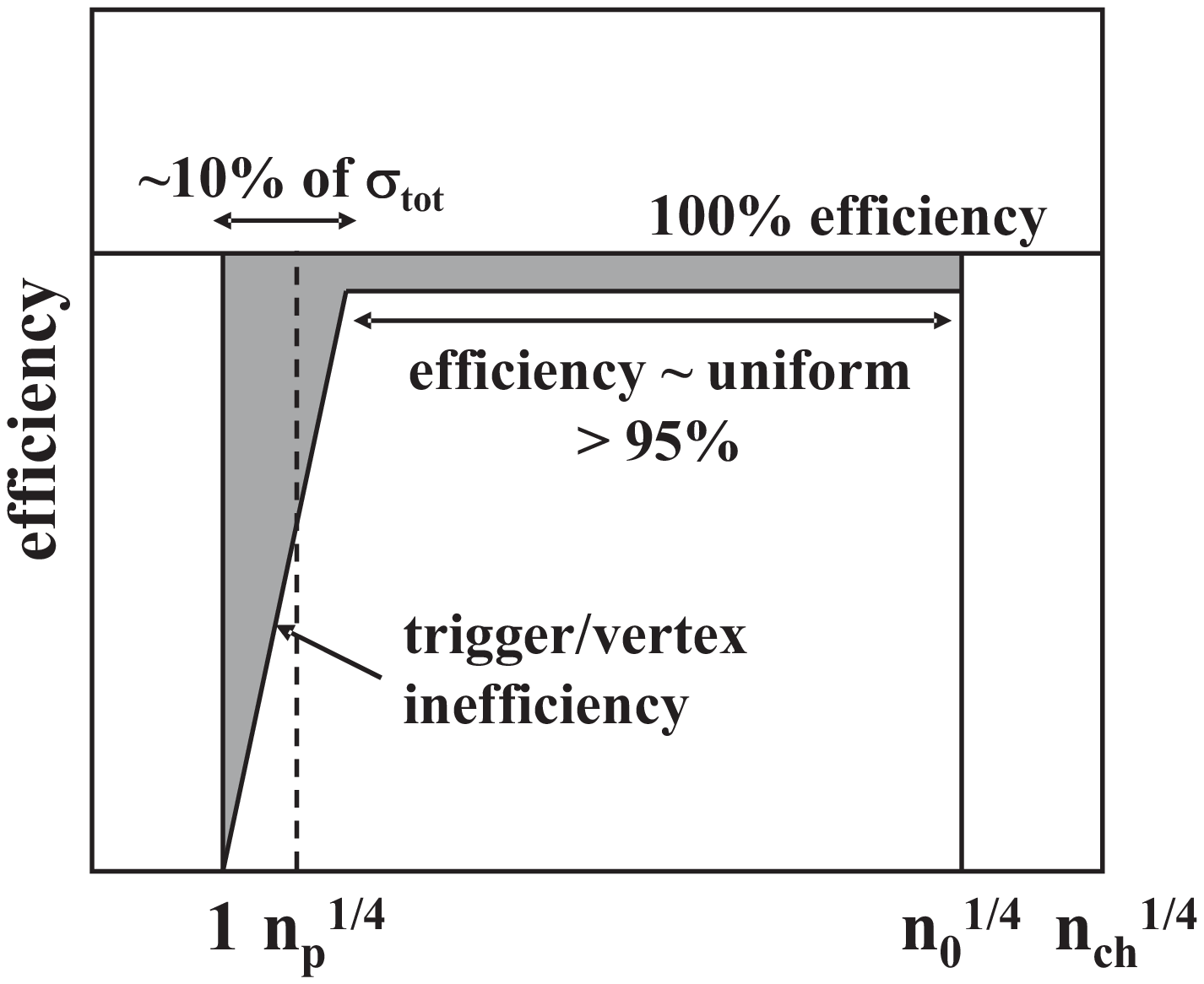}
\includegraphics[width=1.65in,height=1.53in]{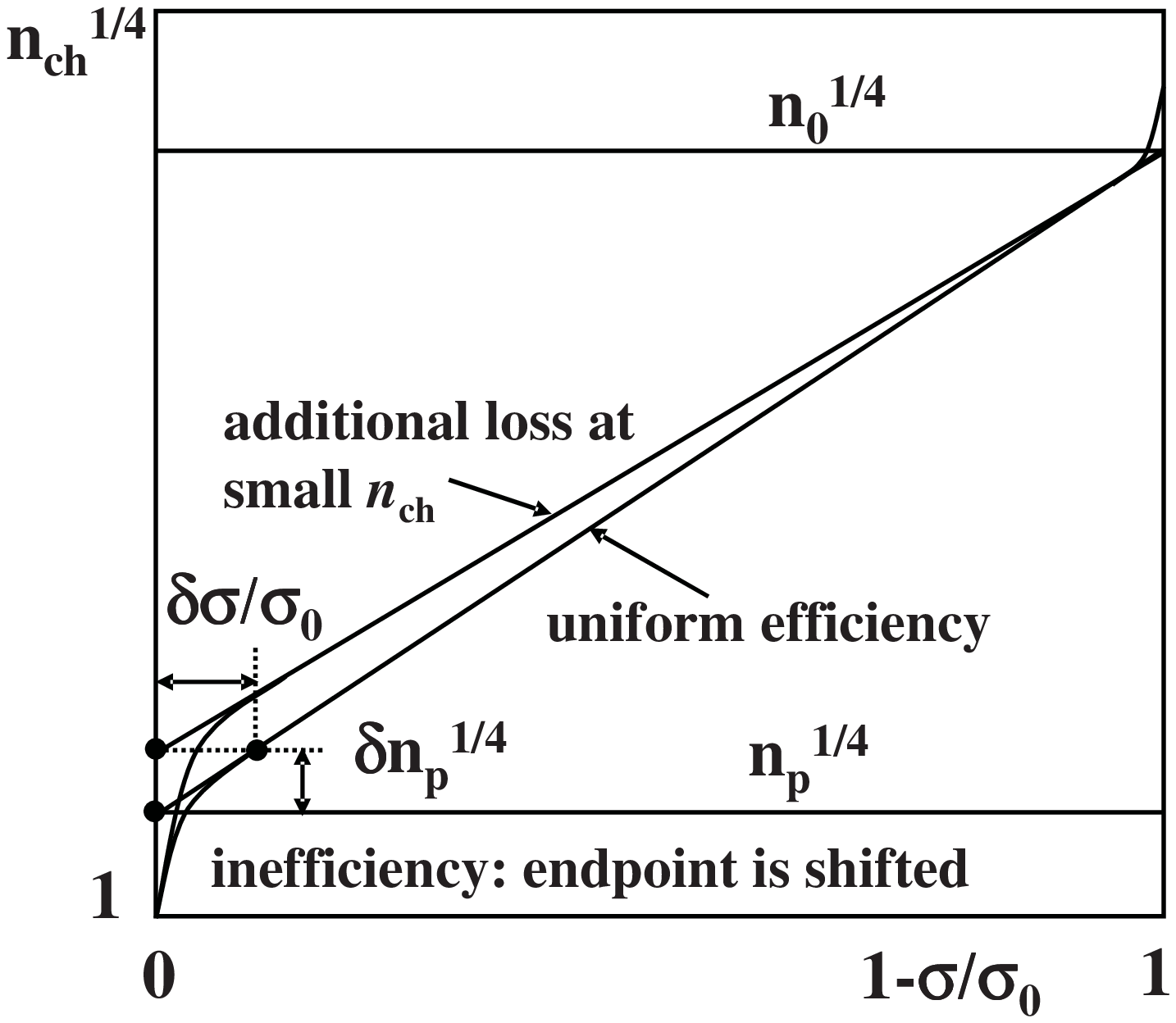}
\caption{Left panel: A cartoon of the power-law differential cross section illustrating typical trigger/vertex inefficiencies. Right panel: A cartoon of running integrals illustrating systematic errors resulting from inefficiencies in the left panel. 
\label{fig10}}
\end{figure}

Trigger/vertex systematic errors are illustrated in Fig.~\ref{fig10} (left panel). The overall event efficiency (trigger plus vertex reconstruction) is typically uniform and close to 100\%, except for peripheral collisions where both efficiencies may be much reduced and strongly varying. The upper endpoint $n_0$ is precisely known ($< 2$\%) from the power-law minimum-bias distribution (but not the conventional method). The lower endpoint $n_p \sim n_{NN}/2$ can be inferred from independent experiments. Those two numbers provide a precise extrapolation reference for the power-law relation between $n_{ch}$ and $\sigma/\sigma_{0}$, as illustrated in Fig.~\ref{fig10} (right panel) .

Assuming a modest centrality error $\delta \sigma / \sigma_0 \sim 1$\%, what is the corresponding relative error of the extrapolated $n^{1/4}_p$? The relation is $\delta n_p/n_p = 4 \delta n_p^{1/4}/n_p^{1/4} = 4 (n_0^{1/4} / n_p^{1/4}  -1) \delta \sigma / \sigma_0 $. For Au-Au at 200 GeV in two units of pseudorapidity $\delta n_p/n_p \simeq  4 ((1050/4)^{1/4} - 1) \times 1 $\% = 12\%. By reverse argument we conclude that if one knows the N-N multiplicity to about 10\% one can limit systematic error in the fractional cross section for data to 1\% for peripheral collisions, another extrapolation constraint. That constraint in turn greatly reduces uncertainty in the Glauber parameters. The most peripheral bin is thereby restored to precision centrality studies. 

\section{Discussion}

To unravel the complex dynamics of heavy ion collisions we must distinguish overlapping contributions to the final-state momentum distribution from initial parton scattering, subsequent in-medium parton dissipation and  fragmentation and bulk-medium dynamics and fragmentation. Such a decomposition requires {\em differential} single-particle spectrum and two-particle correlation analysis precisely related to centrality over the entire centrality range from N-N to central A-A collisions. We require precise knowledge of participant and binary-collision numbers, nucleus overlap geometry and mean participant path length. In this study we have compared conventional and power-law centrality methods for Glauber model simulations, the Hijing Monte Carlo and RHIC data.

\subsection{Advantages of the power-law centrality method}

The power-law method offers the following advantages: 1) precise visual access to the entire minimum-bias distribution for any collision observable,  2) well-defined endpoints $n_p$ and $n_0$ of the minimum-bias $n_{ch}$ (or $p_t$, $E_t$) distribution, with exact correspondence to the endpoints of the Glauber $n_{part}/2$ distribution,  leading to 3) {\em a priori} extrapolation constraints which improve centrality accuracy for peripheral collisions up to 10$\times$,  4) simple and accurate parameterizations of Glauber parameters, 5) flexible definition of centrality binning on $n_{ch}$, and 6) useful extrapolation of the minimum-bias distribution to N-N collisions, even if the distribution is severely distorted by inefficiencies or backgrounds (next subsection). Those features provide first access to the most peripheral 20\% of RHIC A-A collisions.

\subsection{Recovering from measurement distortions}

If the minimum-bias distribution is badly distorted by triggering/vertex inefficiencies or backgrounds the power-law method provides {\em fall-back} determination of collision centrality with good accuracy. We can reconstruct the minimum-bias distribution on $n_{ch}$ from a few pieces of information. Assume that 1) the upper half of the distribution is undistorted  except for smoothly-varying tracking inefficiency, 2) the tracking inefficiency $\epsilon$ for peripheral collisions is known to 10\% and 3) the p-p multiplicity $n_{pp}$ is known to 5\%. From the upper half of the distribution we obtain $n_0$ and the mean slope of the distribution. From the p-p data and the measured efficiency we obtain $n_p = 0.45\, \epsilon\, n_{pp}$ and $n_0^* = 0.45\, \epsilon\, n_{part,0}/2 \cdot n_{pp}$. Those parameters provide the linear participant-scaling reference. The systematic uncertainty in $\nu$ given those conditions is $< 2$\% for peripheral collisions. 

From $n_p$ and $n_0$ we estimate the linear power-law reference. Given the mean slope of the differential distribution from its upper half and Eq.~(\ref{runint}) we can estimate the curved trajectory which lies between the linear limits. We can use either the resulting complete parameterization or the running integral of the data adjusted so that it agrees tangentially with the participant-scaling reference in the peripheral region to determine centrality based on $n_{ch}$ accurate to $< 2$\% over the full centrality range. Examples of such extrapolations and reconstructions are found in Figs.~\ref{fig3}, ~\ref{fig7} and ~\ref{fig13} of this paper.

\subsection{Precision study of particle, $p_t$ and $E_t$ production}

Precise study of particle, $p_t$ and $E_t$ production {\em per participant} relative to centrality is essential to separate initial-state production mechanisms (parton scattering), in-medium parton dissipation, bulk-medium dynamics and final-state hadron production (parton and bulk-medium fragmentation). Collision dynamics should be confronted {\em separately} for total $n_{ch}$, $p_t$ and $E_t$ in a large acceptance {\em vs} path length $\nu$. In contrast, $\langle p_t \rangle$ or $\langle e_t \rangle$ (lower curves in Fig.~\ref{fig12} -- right panel) measure $p_t$ and $E_t$ {\em per final-state hadron}, and are thus {strongly modified by the final-state hadronization process}, mixing initial-state and final-state collision mechanisms. 

We have shown in Figs.~\ref{fig11}, \ref{fig12} and \ref{fig13} that the power-law centrality method reveals new details of the collision process at the few-percent level and reverses some apparent contradictions. We have related several manifestations of the two-component model which provides a simple linear reference for production mechanisms. Significant deviations from that model, obtained for the first time with the power-law method, offer further details of the A-A collision process and bulk-medium properties. The per-participant distributions on $\nu$ provide the means to look beyond hadronization into the dynamics of the pre-hadronic QCD medium.

\section{Summary}

We have introduced a new technique for determining heavy ion collision centrality based on an observed {\em power-law} trend in the minimum-bias distribution on particle multiplicity $n_{ch}$ of the form $n_{ch}^{-3/4}$. The power-law trend implies that transformation to $n_{ch}^{1/4}$ (power-law format) should result in a nearly uniform distribution with {\em physics-related} variations confined to about $\sim20$\% of the mean value. The linear power-law plotting format provides precise visual access to distribution structure. The clearly-identified endpoints (half-maximum points) of the minimum-bias data distribution correspond to the endpoints of the participant-nucleon and binary-collision distributions, providing a precise relation between data and Glauber simulations used to connect collision geometry to measured collision observables. 

We have confirmed that the power-law plotting format is applicable to Glauber $n_{bin}$ and $n_{part}$ minimum-bias distributions as well as to data $n_{ch}$, $p_t$ and $E_t$ distributions. We find that the minimum-bias distribution on $n_{part}$ is almost exactly $\propto n_{part}^{-3/4}$, explaining the similar trend in data, and the distribution on $n_{bin}$ is approximately $\propto n_{bin}^{-5/6}$. Distributions on $n_{part}^{1/4}$ and $n_{bin}^{1/6}$ are therefore nearly uniform, providing simple and precise linear representations of the Glauber parameters {\em vs} fractional cross section. 

We have shown that application of power-law techniques to A-A centrality determination can reduce systematic errors for peripheral collisions to the percent level, even when measurements are severely distorted by inefficiencies. The sharp reduction in systematic uncertainties results from application of extrapolation constraints accessible only in the power-law context. 

We have applied the new centrality techniques and power-law parameterizations of the Glauber parameters to studies of particle, $p_t$ and $E_t$ production in Hijing and RHIC collisions. We have shown that the power-law method can reduce systematic errors in $E_t$ production of up to 50\% to systematic uncertainties of about 5\%. We have also included three appendices which review particle-production algebra and numerical integration techniques required to optimize the accuracy of collision geometry in the power-law context.

TAT appreciates helpful discussions with J. C. Dunlop (Brookhaven National Laboratories) and R. L. Ray (University of Texas, Austin) which led to Appendix~\ref{intfluct}.
This work was supported in part by the Office of Science of the U.S. DoE under grant DE-FG03-97ER41020.

\appendix

\section{Particle Production} \label{partprod2}

According to a two-component model of particle production ($n_{ch}$, as well as $p_t$  and $E_t$ production) centrality trends may follow a mixture of participant and binary-collision scaling~\cite{nardi}. The two-component model is governed by a single parameter $x$ and expressed in the form  
\bea \label{twocomp}
2n_{ch}/n_{part} = n_{NN} (1 + x[\nu - 1]),
\eea  
where $n_{NN}$ is the N-N multiplicity and $\nu$ is the mean participant path length~\cite{nardi}. The model has manifestations in the differential cross section, its running integral and the per-participant-pair particle-production trend. We compare data to a participant-scaling reference in each case. We assume for simplicity that the participant minimum-bias distribution $d\sigma/ d(n_{part}/2)^{1/4}$ is uniform (no sinusoid component). Then, by the chain rule
\bea \label{partcross}
\frac{d\sigma}{dn_{ch}^{1/4}} \hspace{-.05in} &=& \hspace{-.05in} \frac{\sigma_0}{(n_{part,0}/2)^{1/4} - (n_{part,p}/2)^{1/4}  } \hspace{-.03in} \times  \\ \nonumber
&& \hspace{-.03in} \frac{d(n_{part}/2)^{1/4}}{dn_{ch}^{1/4}} ,
\eea
where the first factor is the uniform participant-scaling reference, and the second factor is a Jacobian representing the two-component model which we now calculate. 

Given the two-component relation Eq.~(\ref{twocomp}) we have
\bea \label{runint}
n_{ch}^{1/4}  \approx n_{NN}^{1/4} \left(\frac{n_{part}}{2}\right)^{1/4} \hspace{-.1in} \times \left(1 + x \left[\left(\frac{n_{part}}{2}\right)^{1/3} \hspace{-.1in} - 1 \right]\right)^{1/4} \hspace{-.22in},
\eea
where for this derivation we have used the approximation $\nu \approx (n_{part}/2)^{1/3}$ ({\em cf.} Fig.~\ref{fig6} -- left panel, dash-dot curve) consistent with the approximation $n_{bin} \approx (n_{part}/2)^{4/3}$. The first two factors on the RHS form the participant-scaling reference. Eq. (\ref{runint}) precisely describes centrality-determination plots for corrected data, such as Fig.~\ref{fig8} (right panel).

The required Jacobean factor is obtained from
\bea
\frac{dn_{ch}^{1/4}}{n_{NN}^{1/4}\, d(n_{part}/2)^{1/4}} &\approx&  (1 + x [\nu  - 1])^{1/4} \\ \nonumber
& &\hspace{-.4in} + (n_{part}/2)^{1/4} (1 + x [\nu  - 1])^{-3/4}   x \nu / 3.
\eea
Introducing the Jacobian into Eq.~(\ref{partcross}) we obtain
\bea
\frac{d\sigma}{dn_{ch}^{1/4}}\hspace{-.03in}  &\approx& \hspace{-.03in} \frac{\sigma_0}{n_{NN}^{1/4}\,([n_{part,0}/2]^{1/4} - [n_{part,p}/2]^{1/4})  }
\\ \nonumber
& &  \times \frac{(1 + x [\nu  - 1])^{3/4}}{1 +x (\nu - 1) + x \nu^{5/4}/3} , 
\eea
which for $x > 0$ predicts an approximately linear reduction of the differential cross section on $n_{ch}^{1/4}$ extending from $n_{NN}^{1/4}$ to $n_0^{1/4}$. Assuming $x \sim 0.08$ for 130 GeV RHIC Au-Au data we predict a fractional reduction for central collisions of about 20\% relative to the uniform participant-scaling reference (first factor on the RHS). In Fig.~\ref{fig3} we indeed observe a slope of $\sim -0.15$ with $\sim$20\% linear reduction over the full centrality range.

We have thus related three examples of power-law centrality determination---the minimum-bias differential distribution, its running integral and the per-participant-pair particle-production trend---with a system of model functions having two parameters: $n_{NN}$ and $x$. Experimentally, the measured combination of $n_{NN}$ and $n_0$ determines $x$. For instance, at 200 GeV $n_0 \approx 191\, n_{NN} (1+ 4.95\, x)$. The $x$ value so inferred must be consistent with the negative slope of the power-law minimum-bias distribution, the curvature of its running integral and the positive slope of the particle-production trend. The last format, through the running integral, reduces the short-wavelength statistical noise on the minimum-bias distribution at the expense of introducing long-wavelength noise from uncertainty in the Glauber $n_{part}/2$ parameterization. The same relationships are true for $p_t$ and $E_t$, as demonstrated in Figs.~\ref{fig12} and~\ref{fig13} (although $x$ is unique for each case).

\section{Numerical Integration} \label{runbin}

A major source of centrality error is relating Glauber parameters to measured quantities through running integrals of the fractional cross section. We want to reduce those errors by improving the running-integration methods. To that end we review some details of binning and numerical integration.

\subsection{Binning} \label{binning}

Confusion may arise in comparing the histogram of a density on a binned continuous variable with a distribution on a discrete (integer) variable. A histogram bin entry represents the integral of a density over the bin. When divided by a bin width, the entry estimates a {\em sample} of the density within the bin. If a precise correspondence is sought between distributions on continuous and discrete variables (e.g., $E_t$ and $n_{part}/2$) or two discrete variables (e.g., $n_{ch}$ and $n_{part}/2$), optimized binning and integration definitions are required. 

\subsection{Power-law transformation} \label{power-law}

The elements of a minimum-bias distribution are event counts $N_i$, where $i$ is a bin index on continuous variables $p_t$ or $E_t$ or labels values of discrete variables $n_{ch}$, $n_{part}/2$ or $n_{bin}$.  For bin widths $\delta x_i$ on $x$ the minimum-bias {\em density} is estimated by $dN_i/dx = N_i / \delta x_i$. The power-law form is obtained by the transformation $dN/dx^{1/4} = 4 x^{3/4} dN/dx$. The power-law distribution on multiplicity is $dN/dn_{ch}^{1/4} = 4 n_{ch}^{3/4} dN/dn_{ch} = 4n_{ch}^{3/4}\, N_{n_{ch}}$. The conventional minimum-bias distribution is a non-uniform distribution on a uniform bin system. The power-law form is a nearly-uniform distribution on a non-uniform bin system. The latter makes details of the density distribution and its integration more accessible and facilitates extrapolation. 

\subsection{Rebinning} \label{rebinning}

When transforming a histogram from space $x$ to space $y$ rebinning may be desirable. Assume a distribution $f_i$ ({\em e.g.,} event number) on discrete variable $ n_i$. We want a binned distribution $g_k$ on continuous variable $y$ with uniform bin widths $\delta y$, transformation $y_i = n_i^{1/4}$ and values $g_i = 4n_i^{3/4}\, f_i$. We define uniformly-spaced bin {\em edges} $\hat y_k$ on $y$ with index $k$. For rebinning we step through values $n_i$ and histogram elements $g_i$. Starting with bin $k=1$ on $y$ we sum histogram elements $g_i$ within the $k^{th}$ bin into $G_k$, number of steps on $n$ into $M_k$, and values $n_i$ into $N_k$ (this could also be values $n_i^{1/4}$). We test the $y$ bin edge for each $n_i$: if $n_i^{1/4} < \hat y_k$ we continue the $n$ loop. If not, we advance to $y$ bin $k+1$ and continue the $n$ loop. We increment $k$ until the end of the specified $y$ interval. We then form $G_k / M_k = \bar g_k$ and $N_k / M_k = \bar n_k$ (this could also be $\overline{n^{1/4}}_k$) as uniformly-spaced transformed densities $g_k$ and means $\bar n_k^{1/4}$ (or $\overline{n^{1/4}}_k$) on bin centers $y_k = \hat y_k - \delta y/2$. The minimum-bias distribution $\{f_i\}$ on particle multiplicity $n_{ch}$ is thus rebinned to a `power-law' distribution $\{ g_k\}$ on $n_{ch}^{1/4}$, as in Fig.~\ref{fig2} (left panel).

\subsection{Numerical integration}

To form the Riemann sum of function $f(x)$ over a closed interval $[x_a,x_b]$ of continuous variable $x$ the interval is partitioned into $M$ bins  $i$ of width $\delta x_i$ (not necessarily equal), and sample points $x_i$ are chosen within the bins to sample the density $f(x) \rightarrow f_i = f(x_i)$.  The Riemann sum is $F(\hat x_a,\hat x_b) = F_m = \sum_{i=1}^M \delta x_i\, f_i$, where $\hat x_a$ is the lower edge of the first bin and $\hat x_b$ is the upper edge of the last bin. Depending on the choice of sample positions within the bins the Riemann sum definition is continuously variable from `lower' to `upper' sum. Typically, the $x_i$ are bin centers (`middle' Riemann sum), but we are free to adjust sample positions relative to bin edges to insure mutual compatibility of running integrals. 

The Riemann sum on discrete variable $n$ of distribution $f_i$ at integer values $n_i$ with $i \in [1,M]$ is apparently simple:  $F_M = \sum_{i=1}^M f_i$.  However, if we extend $n$ to a continuous variable with unit-width bins centered on integer values the Riemann sum can be adjusted by shifting the bin edges relative to fixed integer positions (as opposed to shifting sample points within fixed bins). We require that flexibility to insure compatibility of various running integrals defined below. 

\subsection{Running integrals}

To apply the Glauber model precisely we must define the running integral on integer variables (e.g., $n_{part}/2$) consistently with integrals on other integer variables (e.g., $n_{ch}$) or continuous variables (e.g., $p_t$ or $E_t$). To insure compatibility among running integrals on discrete and continuous variables we treat discrete variables as continuous binned variables. The running integral on a continuous variable is $F(\hat x_a,\hat x_m) = F_m = \sum_{i=1}^m \delta x_i\, f_i$, where $\hat x_m$ is the upper edge of the $m^{th}$ bin and $m \in [1,M]$. The running integral of the distribution must be properly related to its independent variable. The corresponding running integral of the independent variable is $\hat x_{m} = \hat x_a +  \sum_{i=1}^m  \delta x_i   $, assuming that integration is between leading and trailing bin edges.

If $f$ on $n$ is transformed to $f^{1/4}$ on $n^{1/4}$ the bin edges are not symmetrically placed about the sample point, and individual bin edges and bin widths must be determined explicitly to minimize discretization errors. We use parameters $u$ and $v$ with $u + v = 1$ to denote bin-edge positions relative to integer positions on $n$. Thus, bin edges relative to $n_i$ are at $n_i+u$ and $n_i-v$. From the bin edges we calculate bin widths $\delta n^{1/4} \equiv   (n+u)^{1/4} - (n-v)^{1/4} $. $(u,v)$ can be adjusted to satisfy extrapolation constraints. E.g., $u = 0.45$ rather than 0.5 due to the NBD distribution skewness and $n_{part,p}/2 = 0 + u = 0.45$. The running integrals are then
\bea
\hat n^{1/4}_{m} &=&  \hat n^{1/4}_{a} + \sum_{i=1}^{m} \delta n^{1/4}_{i} = (n_m + u)^{1/4}  \\ \nonumber
F_m&=& \sum_{i=1}^{m}  f_{i}^{1/4} \cdot   \{(n_i+u)^{1/4} - (n_i-v)^{1/4}\} .
\eea
We plot the running integral and running limit as $F_m$ {\em vs} $\hat n^{1/4}_{m}$ (e.g., Fig.~\ref{fig5b}). It should be clearly indicated whether a plot represents bin entries and bin centers or running-integral sums and corresponding bin edges. Because of the strong nonlinearity of $n^{1/4}$, exact bin-edge positions relative to the integers are very significant for small $n$ and irrelevant for large $n$.

\section{Relating centrality parameters} \label{running}

The underlying premise of centrality determination is that collision observables and Glauber geometry parameters are related statistically by joint probability distributions which are not directly observable. What are accessible are projections of the joint distributions onto their margins. Pairs of variables (measured and Glauber) can then be related by running integrals of the marginal distributions. We now discuss technical details of that procedure.

\subsection{Joint distributions}

The basis for relations between $n_{ch}$, $p_t$ and $E_t$ on the one hand and  $n_{part}/2$ and $n_{bin}$ on the other  is joint density distributions on pairs of variables: a measured quantity and a collision-geometry parameter (e.g., $n_{ch}$ and $n_{part/2}$). The {\em locus of means} (curve describing conditional means) provides the functional relation between two quantities. Joint distributions are not directly observable experimentally, but one projection (marginal distribution on $n_{ch}$, $p_t$ or $E_t$) is measured, and the other (marginal on $n_{part}/2$ or $n_{bin}$) is obtained from a Glauber Monte Carlo simulation. We relate one variable to the other through running integrals of the marginal projections. Each normalized running integral is the fractional cross section in the form $1 - \sigma/\sigma_0$. In the power-law context precise access to the locus of means is provided by the endpoints or half-maximum points of the marginal (minimum-bias)  distributions. 

For  joint distribution $\rho(x,n)$ there are two conditional distributions: $\langle x \rangle_n$ (mean of $x$ for fixed $n$) and $\langle n(x) \rangle $ (mean of $n$ for fixed $x$). We expect $\langle x \rangle_n$ and $\langle n(x) \rangle $ to correspond for intermediate values of $x$ and $n$, but to diverge near the joint distribution endpoints due to `fluctuations' (finite width of the joint distribution). In the power-law context the nearly-linear relation between centrality parameters over most of the joint distribution can be determined precisely by the marginal endpoints determined as the half-maximum points of the marginal distributions. 

\subsection{$n_{part}$ scaling and $n_{ch}$} \label{nparscale}

A key element of A-A centrality determination is the participant-scaling (wounded-nucleon) hypothesis: a projectile nucleon can `participate' in at most one N-N collision, and each such participant pair produces final-state hadrons as in an isolated N-N collision. In that limit the A-A multiplicity `scales' as $n_{ch} \propto n_{part}/2$. Alternatively, in the binary-collision limit A-A multiplicity scales as $n_{ch} \propto n_{bin}$. For strict participant scaling $n_{part}$ cannot have odd values; e.g., there cannot be exactly three participants in an A-A collision. Once a nucleon has participated in an N-N interaction (is `wounded') it cannot produce a third participant in a second encounter, even though the number of binary collisions does increase by one in the binary-collision context. In the participant-scaling limit $n_{part}/2$ should assume only integer values, and in this paper we treat that symbol combination as the basic statistical variable. 

We can combine participant scaling with the power-law form of the minimum-bias distribution to define a precise relation between $n_{ch}$ and $n_{part}/2$ valid to  $< 5$\% for peripheral collisions. In the limiting case $n_{ch} \propto n_{part}/2$ is strictly obeyed (ignoring N-N fluctuations). We employ an {\em extrapolation constraint} at the lower endpoints to provide precise registration of  $n_{ch}$ and $n_{part}/2$, even if the measured minimum-bias distribution on $n_{ch}$ is severely distorted by inefficiencies and backgrounds. If N-N fluctuations are symmetric about the mean then the minimum-bias distributions on $n_{part}/2$ and $n_{ch}$ extrapolate to $n_{part,p}/2 = 1/2$ and $n_p = 1/2 \cdot n_{NN}$ respectively.   The upper endpoints of the participant-scaling reference go to $n_{part,0}/2$ and $n_0 = n_{part,0}/2 \cdot n_{NN}$, where $n_{part,0}/2 \sim 191$ at $\sqrt{s} =$ 200 GeV. In the next section we justify those endpoint values with simulations and show how to achieve precise correspondence for real N-N fluctuations with nonzero skewness.

\subsection{Running integrals and N-N fluctuations} \label{intfluct}

\begin{figure}[h]
\includegraphics[width=3.3in,height=3.3in]{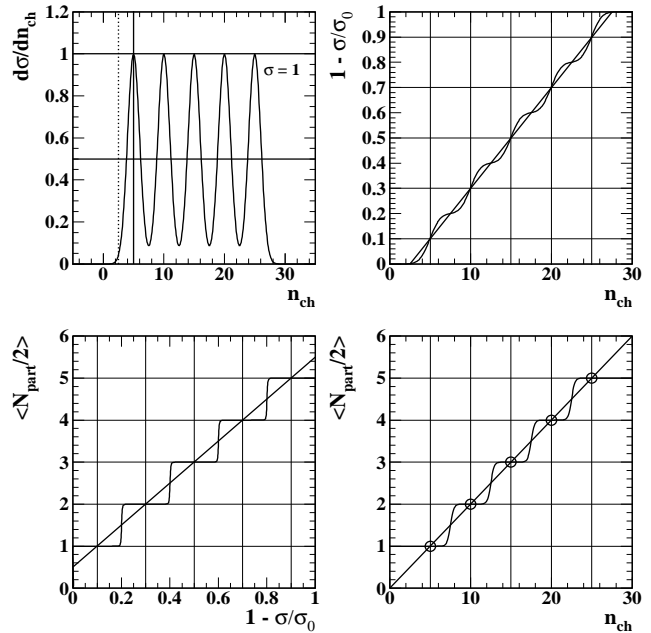}
\caption{Upper-left panel: Minimum-bias distribution on $n_{ch}$, with narrow gaussian N-N distribution. Upper-right panel: running integral of the minimum-bias distribution on $n_{ch}$ to obtain the fractional cross section. Lower-left panel: Conditional distribution of $\langle n_{part}/2 \rangle$ on fractional cross section. Lower-right panel:  Conditional distribution of $\langle n_{part}/2 \rangle$ on $n_{ch}$. The open circles denote the complementary conditional distribution  $\langle n_{ch} \rangle$ on $n_{part}/2$.    
\label{intquad1}}
\end{figure}

\begin{figure}[h]
\includegraphics[width=3.3in,height=3.3in]{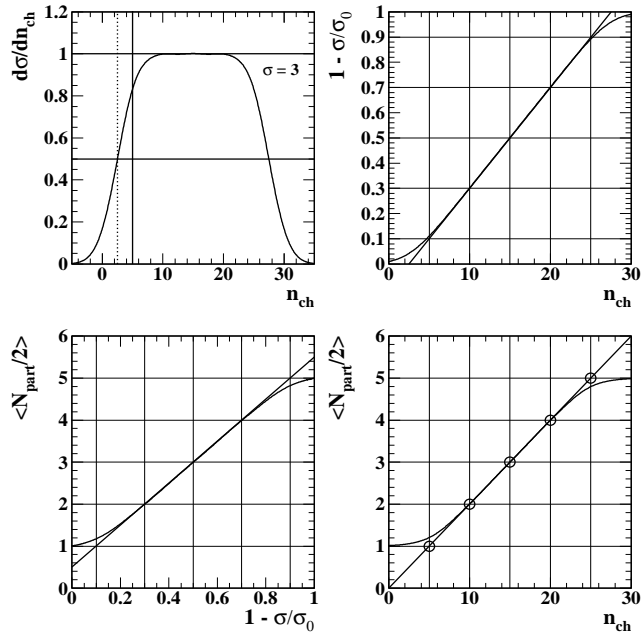}
\caption{The same plots as in Fig.~\ref{intquad1}, but with broader N-N multiplicity distribution to illustrate asymptotic behavior near the endpoints for RHIC data.
\label{intquad2}}
\end{figure}

N-N fluctuations can change the position of endpoint $n_p$ relative to $n_{NN}$, or equivalently the value of $n_{part,p}$.  To understand the systematics we model the joint distribution  on $(n_{part},n_{ch})$.  We assume strict participant scaling. We also assume that the differential cross section is constant on $n_{part}/2$ rather than $(n_{part}/2)^{1/4}$. The projection onto $n_{part}/2$ is a set of delta functions on the integers. The projection onto $n_{ch}$ depends on N-N fluctuations, as shown in Figs.~\ref{intquad1} and \ref{intquad2}. We relate $n_{ch}$ to $n_{part}/2$ through running integrals of the projections. 

In Fig.~\ref{intquad1} (upper-left panel) we show the projected density on $n_{ch}$ for a narrow gaussian N-N multiplicity distribution and 1 -- 5 participant pairs, assuming $n_{NN} = 5$ and $\sigma = 1 < \sqrt{n_{NN}}$. In the upper-right panel we show the running integral of fractional cross section $1 - \sigma/\sigma_0$ on $n_{ch}$. In the lower-left panel we show the conditional distribution $\langle n_{part}/2\rangle$ on $\sigma/\sigma_0$ (different from the running integral on $n_{part}/2$), and in the lower-right panel we show the conditional distribution $\langle n_{part}/2\rangle$ on $n_{ch}$. The open circles indicate the complementary discrete conditional distribution $\langle n_{ch} \rangle$ on $n_{part}/2$. The two are consistent by construction. The lower-right panel reflects a `chain-rule' relating $\langle n_{part}/2\rangle$ to $n_{ch}$ through the lower-left and upper-right panels.

In Fig.~\ref{intquad2} we show the same set of plots for  $\sigma = 3 > \sqrt{n_{NN}}$. The smooth variations allow us to examine the structure at the endpoints. In the upper-left panel we see that the lower half-maximum (dotted line) falls at  $n_p = n_{NN}/2 = 2.5$  ($n_{part,p} = 1/2$). That endpoint is stable over a range of N-N widths for {\em symmetric} N-N fluctuation distributions. In the upper-right panel the running integral endpoints (extrapolated values) correspond to the half-maximum points of the differential distribution. In the lower-left panel the lower endpoint is $\langle n_{part,p}/2 \rangle = 1/2$ as expected. 

In the lower-right panel the locus of means for the two complementary conditional distributions (solid curve and dots) agrees over the central region, as determined by the match of endpoint values in the lower-left and upper-right panels. For real systems we must adjust the Riemann sum definitions so that (for instance) the locus-of-means endpoints on $n_{ch}$ and $n_{part}/2$ coincide. Also, we must define and integrate $n_{part}/2$ and $n_{bin}$ so that both lower limits are the same to insure that $\nu$ goes to 1 in the limit.

The minimum-bias distribution for real data is based on an N-N multiplicity distribution with nonzero skewness, commonly modeled by the negative binomial distribution (NBD). The NBD is defined by
\bea
P(n;k,p) = \frac{(k+n-1)!}{(k-1)!(n)!}(p-1)^k p^n,
\eea
with fixed $k$. For comparison, the binomial distribution with the same notation is
\bea
P(n;N,p) = \frac{(k+n)!}{(k)!(n)!}(p-1)^k p^n
\eea
with fixed $N = k + n$. The NBD has mean $\mu = kp/(1-p)$ and variance $\sigma^2 = \mu/(1-p)$, whereas the binomial has $\mu = N p$ and variance $\sigma^2 = \mu (1-p)$. Both go to a Poisson distribution with $\sigma^2 = \mu$ if $N, k \rightarrow \infty$ and $p \rightarrow 0$ with fixed $\mu$. For the NBD $(\sigma^2_{NBD} - \sigma^2_{Poisson}) / \mu = \mu / k = p / (1-p)$ is a measure of normalized variance (fluctuation) excess relative to Poisson due to correlations. For N-N collisions in the STAR TPC acceptance $\mu \sim 5$ and $k \sim 10$, with $p \sim 1/3$.

We want to understand the effect of skewness and width changes on the value of $n_{part,p}$. The NBD skewness depends on its control parameter $k$. If $\mu/k \rightarrow 0$ the NBD goes to a Poisson distribution. For larger $\mu / k$ the NBD is increasingly skewed and its width increases. The lower half-maximum point on the differential cross section then moves below $n_{NN}/2$ and the participant-pair-number running integral definition must change to accommodate. 

\begin{figure}[h]
\includegraphics[width=1.65in,height=1.65in]{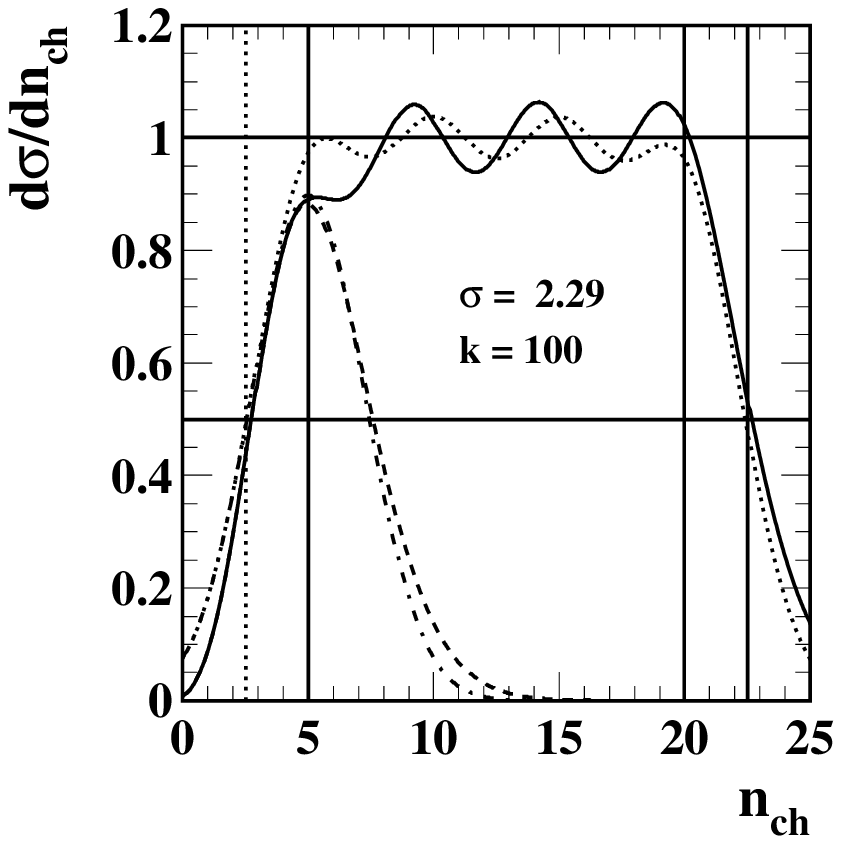}
\includegraphics[width=1.65in,height=1.65in]{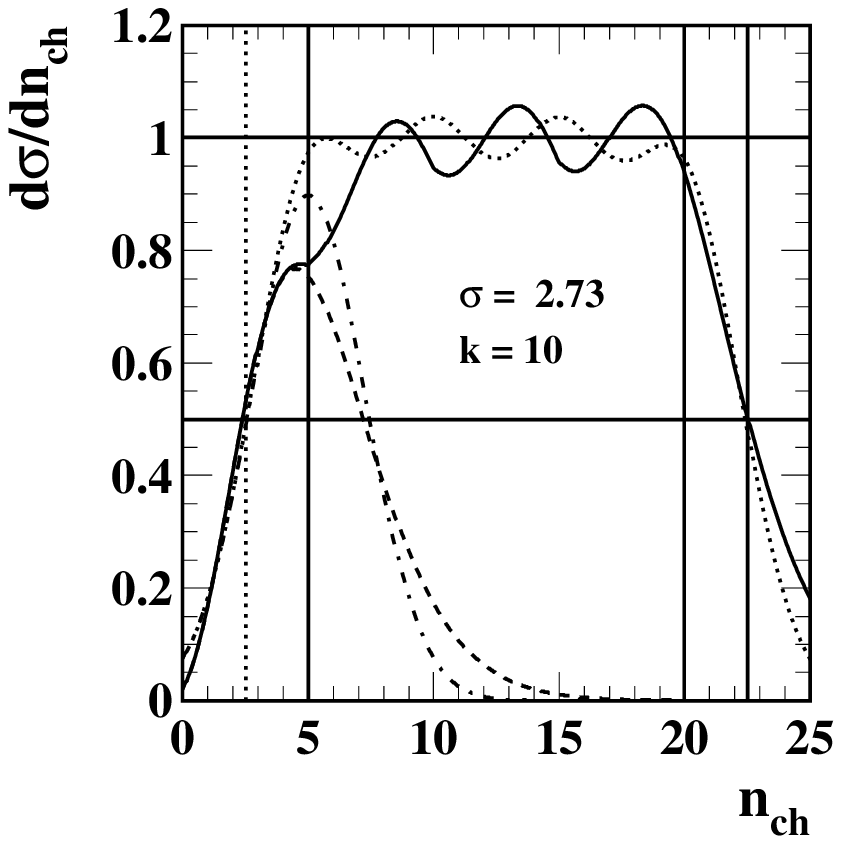}
\caption{Model of a minimum-bias distribution on $n_{ch}$ with NBD N-N distribution, showing the changing position of endpoint $n_{p} \sim 2.5$ relative to $n_{NN} = 5$ as the NBD correlation parameter $k$ is varied. For RHIC data ($k \sim 10$) $n_p \sim 0.45 \, n_{NN}$ is a good approximation. 
\label{glaub13-3a}}
\end{figure}

In Fig.~\ref{glaub13-3a} we compare minimum-bias distributions based on NBD and gaussian distributions. In  the left panel we show a minimum-bias distribution (solid curve) with the individual N-N distributions modeled by an NBD with $k = 100$ and $n_{NN} = 5$ (dashed curve) . For comparison the dotted and dash-dot curves are based on a gaussian N-N model, as in the previous figure but with $\sigma = \sqrt{5} = 2.25$. The value of $n_p$ for the NBD is slightly larger than the $ n_{NN}/2 = 2.5$ for the gaussian. The NBD with $k = 100$ is nearly Poisson, with $\sigma$ only slightly larger than $\sqrt{\mu}$.

In Fig.~\ref{glaub13-3a} (right panel) we show similar distributions but with $k = 10$ corresponding to data. Due to the increased skewness and width of the NBD with the smaller $k$ the value of $n_p$ is shifted below $n_{NN} / 2$ to $n_p \approx 0.45\, n_{NN}$. We conclude that $n_{part,p}/2 = 0.45\pm0.05$ is a good match to RHIC A-A data. We therefore shift bin edges on $n_{part}/2$ down slightly relative to the middle Riemann sum. Equivalently, we adjust the $n_{part,p}/2$ value in the parameterization of Eq.~(\ref{npart}) and adopt $n_p \approx 0.45\, n_{NN}$ as the extrapolation constraint for minimum-bias distributions on $n_{ch}$. The 10\% relative uncertainty in $n_{part,p}/2$ propagates to a mere 0.7\% uncertainty in $\sigma/\sigma_0$ for peripheral collisions. However, it corresponds in particle, $p_t$ or $E_t$ production plots such as Fig.~\ref{fig13} to a 5\% uncertainty localized near $\nu = 2$.


\end{document}